\documentclass[
showpacs,
preprintnumbers,
reprint, 
amsmath,amssymb,
10pt
,letterpaper
, aip, jcp
]
{revtex4-1}

\usepackage{graphicx}
\usepackage{dcolumn}
\usepackage{bm}

\usepackage[english]{babel}

\newcommand{\vect}[1]{{\mathbf{#1}}} 
\newcommand{\vgr}[1]{{\bm{#1}}} 
\newcommand{\uvec}[1]{{\mathbf{\hat{#1}}}} 

\newcommand{\gbody}{\bar{g}^{(2)}}

\newcommand{\rh}[1]{\rho^{(#1)}}

\newcommand{\curr}[1]{\vgr{\mathcal{J}}_{\!#1}} 

\newcommand{\rev}[1]{{#1}}

\begin{document}


\title{
Particle-scale statistical theory for hydrodynamically induced polar ordering in microswimmer suspensions
}

\author{Christian Hoell}
\email{christian.hoell@uni-duesseldorf.de}

\author{Hartmut L\"owen}

\author{Andreas M.~Menzel}
\email{menzel@hhu.de}
\affiliation{Institut f\"ur Theoretische Physik II: Weiche Materie,
Heinrich-Heine-Universit\"at D\"usseldorf,
40225 D\"usseldorf, Germany.}

\date{\today}

\begin{abstract}
Previous particle-based computer simulations have revealed a significantly more pronounced tendency of spontaneous global polar ordering in puller (contractile) microswimmer suspensions than in pusher (extensile) suspensions. We here evaluate a microscopic statistical theory to investigate the emergence of such order through a linear instability of the disordered state. For this purpose, input concerning the orientation-dependent pair-distribution function is needed, and we discuss corresponding approaches, particularly a heuristic variant of the Percus test-particle method applied to active systems. Our theory identifies an inherent evolution of polar order in planar systems of puller microswimmers\rev{,} if mutual alignment due to hydrodynamic interactions overcomes the thermal dealignment by rotational diffusion. In the theory, the cause of orientational ordering can be traced back to the actively induced hydrodynamic rotation--translation coupling between the swimmers. Conversely\rev{,} disordered pusher suspensions remain linearly stable against homogeneous polar orientational ordering. We expect that our results can be confirmed in experiments on (semi-)dilute active microswimmer suspensions, based, for instance, on biological pusher- and puller-type swimmers. 
\end{abstract}

\maketitle


\section{Introduction}

Microswimmers \cite{purcell1977life, howse2007self, lauga2009hydrodynamics, elgeti2015physics,zottl2016emergent, bechinger2016active} --- both biological \cite{eisenbach2006sperm,berg2008coli,polin2009chlamydomonas, mussler2013effective,goldstein2015green}  and artificial \cite{paxton2004catalytic,buttinoni2012active,walther2013janus,samin2015self} --- 
have been studied widely and 
can be considered as an archetype of active soft matter.\cite{ramaswamy2010mechanics, marchetti2013hydrodynamics, menzel2015tuned} 
Since these self-propelled particles are inherently in non-equilibrium with their surroundings,
their study has led to rather unexpected findings,
e.g., motility-induced phase separation,\cite{cates2013active,buttinoni2013dynamical,bialke2013microscopic,bialke2015active,cates2015motility,wittkowski2017nonequilibrium,solon2018generalized_NJP,digregorio2018full} 
laning,\cite{wensink2012emergent,menzel2013unidirectional,kogler2015lane,romanczuk2016emergent,menzel2016way}
various kinds of ``taxis'' \cite{dusenbery2009living} by implicit steering, \cite{kessler1985hydrodynamic,durham2009disruption,hagen2014gravitaxis,lozano2016phototaxis,campbell2017helical,liebchen2018viscotaxis}
and bacterial turbulence. \cite{wensink2012meso,wensink2012emergent,sokolov2012physical,dunkel2013fluid,kaiser2014transport, slomka2015generalized}   
Establishing a physical description of the observed collective phenomena calls for the development of new methods in statistical physics. \cite{cates2012diffusive,ten2015can,yan2015force,yan2015swim, takatori2014swim, solon2015pressure, smallenburg2015swim, liluashvili2017mode}
Furthermore, there is a huge amount of biological and medical problems
for which the knowledge about microswimmers and their physical behavior is key, \cite{engstler2007hydrodynamic,kessler1985hydrodynamic,durham2009disruption,nelson2010microrobots,wang2012nano,patra2013intelligent,xi2013rolled,abdelmohsen2014micro}
warranting strong research interest in the topic.

Approaching the scientific field of microswimmers as an extension of the study of colloidal suspensions \cite{Dhont}
allows both experimentalists and theoreticians to carry over methods and ideas.
An important example is hydrodynamics: 
microswimmers typically operate in low-Reynolds-number regimes.\cite{purcell1977life}
In this context a whole apparatus of physical theory \cite{Happel_Brenner, Dhont, Kim_Karrila} is at hand
as a toolkit for, e.g., the investigation of hydrodynamic interactions between swimmers and the influence of these interactions on the collective behavior of microswimmer suspensions.

As a consequence of the swimming at low Reynolds numbers, no net force may be exerted by a model microswimmer on its environment. \cite{purcell1977life,ten2015can}
To the lowest order, the induced flow field of a typical swimmer in general can thus be described as generated by a force dipole 
(we here disregard ``neutral-type'' swimmers with a vanishing \rev{averaged} force-dipole contribution to the flow field like, 
e.g., the famous Najafi-Golestanian three-spheres swimmer \cite{najafi2004simple,zargar2009three,daddi2018swimming,daddi2018state}).
Depending on the orientation of the forces (outwards / inwards) of that dipole,
one can distinguish ``pusher'' (also called extensile) microswimmers --- for which fluid is pushed outwards along the axis of motion and sucked in from the transverse axes ---
and ``puller'' (also termed contractile) microswimmers --- for which the inverse is true.\cite{underhill2008diffusion, baskaran2009statistical}
Since the direction of swimming is given by the orientation of the swimmer, interactions affecting the rotational degrees of freedom are of utmost interest.

A breakthrough in the study of orientational self-organization of self-propelled particles has been the Vicsek model, introducing simple effective local alignment rules. 
They can lead to emergent long-range orientational order in these active systems, even in two spatial dimensions. \cite{vicsek1995novel,toner1995long,farrell2012pattern,vicsek2012collective,liebchen2017collective,levis2018micro}
Such an effective alignment mechanism can be interpreted either as being social in nature, e.g., when applied to flocks of birds, \cite{vicsek1995novel,toner1995long,vicsek2012collective} or as a coarse-grained model representing underlying physical interactions, e.g., steric alignment interactions.\cite{ginelli2010large} 
In the present work, we focus on the question, to which extent hydrodynamic interactions can provide sufficient alignment to result in polarly ordered collective motion.

Previously, corresponding computer simulations have found that indeed hydrodynamic interactions between microswimmers can lead to collective alignment
in pure puller microswimmer suspensions, \cite{alarcon2013spontaneous}
also when doped with pusher microswimmers. \cite{pessot2018}
Typically, the degree of observed orientational order in pure pusher suspensions is notably lower.\cite{alarcon2013spontaneous,pessot2018}
In the current work we analyze a microscopic statistical theory to understand reasons for these differences in polar ordering observed for pushers and pullers.
For this purpose we extend our previously developed dynamical density functional theory (DDFT) of microswimmers,\cite{menzel2016dynamical, hoell2017dynamical} built on the force-dipole-based minimal swimmer model introduced in Refs.~\onlinecite{menzel2016dynamical, hoell2017dynamical, pessot2018}.

A brief recapitulation of the theoretical background follows in Sec.~\ref{sec:theory}.
The theory is then applied to a \mbox{(semi-)dilute} swimmer configuration confined to a plane in Sec.~\ref{sec:application}. 
Next, to theoretically analyze the emergence of collective polar alignment from hydrodynamic interactions, some microscopic details of the (orientation-dependent) pair distribution function are needed as an input.
A reasonable approximation for this pair distribution function is discussed in Sec.~\ref{sec:pdf_approx}. 
As the central step, a linear stability analysis probing the emergence of collective alignment out of the isotropic disordered state is performed in Sec.~\ref{sec:linear}. 
There, indeed we find that hydrodynamic interactions can induce polar ordering in \mbox{(semi-)dilute} suspensions of sufficiently-strong puller microswimmers. 
In contrast to that, linear stability of disorder is found for corresponding spatially homogeneous pusher suspensions.
Finally, a short conclusion and outlook are given in Sec.~\ref{sec:conclusions}.

\section{Theory}
\label{sec:theory}

As just mentioned, this section repeats the central parts of the statistical theory of microswimmers developed in our previous works.\cite{menzel2016dynamical, hoell2017dynamical} 
At the end of the section, a dynamical equation for the one-swimmer density (as defined below) is listed. It is the starting point for our investigation of possibly emerging polar ordering in planar (semi-)dilute microswimmer configurations in Secs.~\ref{sec:application}--\ref{sec:linear}.

\begin{figure}
\vspace{-30pt}
\includegraphics{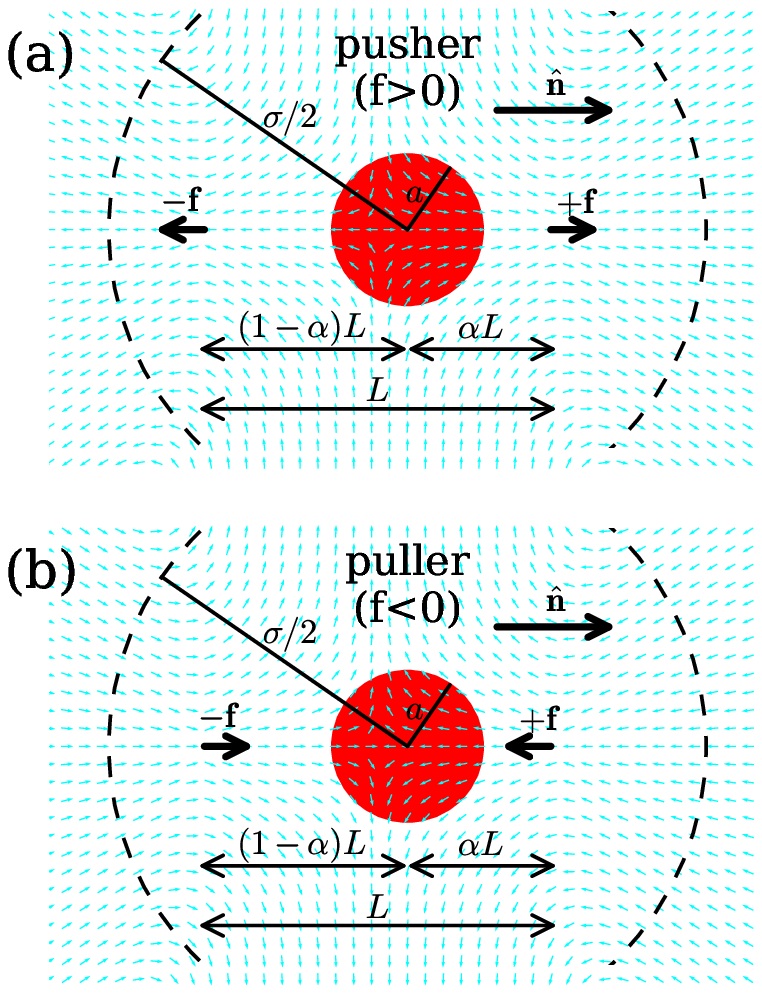}
\vspace{-30pt}
\caption{Minimal microswimmer model, as introduced in Ref.~\onlinecite{menzel2016dynamical}. 
A sphere of radius $a$ constitutes a no-slip boundary for the flow of the surrounding fluid and represents the hydrodynamic swimmer body. 
Two force centers exerting opposite forces $\pm\vect{f}=\pm f\uvec{n}$ of equal magnitude on the fluid are placed nearby in an axially symmetric configuration.
They generate the flow indicated by the small arrows, which propels the swimmer.
This force-sphere combination is rigidly kept in its internal (body-frame) configuration.
(a) For $f>0$, a pusher microswimmer is created, while (b) a puller microswimmer results for $f<0$. 
Other swimmers are exposed to the flow, too, but are kept at distance by a repulsive interaction potential of characteristic range $\sigma$. 
The resulting effective steric \rev{extension} of the swimmer is indicated by the \rev{dashed} line.}
\label{fig:model}
\end{figure}

We consider a suspension of $N$ (identical) axially symmetric microswimmers in a volume $V$. 
Inertial effects are neglected in the investigated low-Reynolds-number regime.
The state of each swimmer $i=1,\dots,N$ is characterized by a phase space coordinate $\vect{X}_i = (\vect{r}_i, \uvec{n}_i)$ that comprises its spatial position $\vect{r}_i$ and its orientation, described by the unit vector $\uvec{n}_i$.
We recur to the minimal swimmer model introduced in Ref.~\onlinecite{menzel2016dynamical}, see Fig.~\ref{fig:model}. 

There, two opposing force centers, exerting forces $\pm\vect{f} := \pm f \uvec{n}$ on the fluid, rigidly move and rotate together with a spherical swimmer body of hydrodynamic radius $a$. 
In terms of the swimmer coordinates, the force centers are located at positions $\vect{r}_{i}^+ :=\vect{r}_i + \alpha L \uvec{n}$ and $\vect{r}_{i}^-:= \vect{r}_i - (1-\alpha) L \uvec{n}$, respectively, with $a / L < \alpha \leq1/2$ a positive number and $L$ the fixed distance between the two force centers.
The rigid spherical swimmer body of no-slip surface condition is located at position $\vect{r}_i$ in the generated flow of the surrounding fluid.
This configuration of the sphere and the two force centers is treated as a rigid entity that translates and rotates as one.
For $\alpha \neq 1/2$, net self-propulsion in the direction of $\mathrm{sign}(f) \uvec{n}$ results.
Accordingly, a pusher (puller) microswimmer \cite{baskaran2009statistical} is  constructed for $f>0$ ($f<0$).
Furthermore, a steric interaction potential between different swimmers with sufficiently large effective diameter $\sigma$ is introduced to counteract unphysical overlap.
By construction, no net force and no net torque are exerted by the swimmer on the fluid, a necessary condition for microswimmers.\cite{purcell1977life,ten2015can}

In the following, a statistical description of the microswimmer suspension is employed. \rev{We} start our approach from the (time-dependent) microstate probability density $P=P(\vect{X}^N,t)$ to find the system in microstate $\vect{X}^N$ at time t, with $\vect{X}^N=\{\vect{X}_1, ..., \vect{X}_N\}$. 
For our overdamped low-Reynolds-number system,\cite{purcell1977life, Dhont} the dynamical evolution of $P$ is described by the many-body Smoluchowski equation 
\begin{equation}
 \frac{\partial P}{\partial t} = - \sum\limits_{i=1}^N \Big[ \nabla_{\vect{r}_i} \cdot (\vect{v}_i P) + \left(\uvec{n}_i \times \nabla_{\uvec{n}_i}\right) \cdot (\vgr{\omega}_i P) \Big],
\label{Smoluchowski}
\end{equation}
where $\vect{v}_i$ is the velocity of swimmer $i$ and $\vgr{\omega}_i$ is its angular velocity, which generally both depend on the configuration $\vect{X}^N$ of the system.

We only take into account pairwise additive hydrodynamic interactions between the swimmers on the Rotne-Prager level.\cite{Dhont}
Neglecting many-body hydrodynamic interactions is a good approximation at low to intermediate densities \cite{beenakker1983self,beenakker1984diffusion,nagele1996dynamics, nagele1997long, banchio2006many} as regarded here. Thus, in the discrete particle picture, $\vect{v}_i$ and $\vgr{\omega}_i$ of swimmer $i$ follow from the forces $\vect{F}_j$ and torques $\vect{T}_j$ acting on all swimmers $j$ via \cite{menzel2016dynamical, hoell2017dynamical}
\begin{align}
\left[
\begin{array}{c}
 \vect{v}_i \\[0.1cm]
 \vgr{\omega}_i
\end{array}
\right]
=&
    \sum_{j=1}^{N}
    \Bigg(
\left[
    \begin{array}{cc}
    \vgr{\mu}^{\mathrm{tt}}_{ij} & \vgr{\mu}^{\mathrm{tr}}_{ij}\\[0.1cm]
    \vgr{\mu}^{\mathrm{rt}}_{ij} & \vgr{\mu}^{\mathrm{rr}}_{ij}    
    \end{array}
\right]
    \cdot
\left[
    \begin{array}{c}
        \vect{F}_j \\[0.1cm]
        \vect{T}_j
    \end{array}
\right] \notag \\
    &+
\left[
    \begin{array}{cc}
    \vgr{\Lambda}^{\mathrm{tt}}_{ij} & \vect{0}\\[0.1cm]
    \vgr{\Lambda}^{\mathrm{rt}}_{ij} & \vect{0}   
    \end{array}
\right]
    \cdot
\left[
    \begin{array}{c}
        f \uvec{n}_j \\[0.1cm]
        \vect{0}
    \end{array}
\right]
    \Bigg),
  \label{mobility}
\end{align}
$i=1,\dots,N$,
exploiting the linearity of the underlying Stokes equation in the low-Reynolds-number regime.\cite{Dhont}
The viscosity $\eta$ of the background fluid is assumed to be constant and the well-known hydrodynamic mobility expressions for passive rigid spheres on the Rotne-Prager level \cite{Rotne_1969_JCP, reichert2004hydrodynamic} are used.
This way, the self mobilities are given by
\begin{equation}
 \bm{\mu}^\mathrm{tt}_{ii} = \mu^\mathrm{t} \bm{1}, \quad \bm{\mu}^\mathrm{rr}_{ii} = \mu^\mathrm{r} \bm{1}, \quad \bm{\mu}^\mathrm{tr}_{ii} =  \bm{\mu}^\mathrm{rt}_{ii} = \bm{0},
\end{equation}
with $\bm{1}$ the identity matrix and 
\begin{equation}
 \mu^\mathrm{t}=1/(6\pi\eta a), \quad \mu^\mathrm{r}=1/(8\pi\eta a^3),
\end{equation}
while the pair mobilities ($j\neq i$) read
\begin{align}
\bm{\mu}_{ij}^{\mathrm{tt}}=&\mu^\mathrm{t}\bigg(\frac{3a}{4r_{ij}}\Big({\bf {1}}+{{\mathbf{\hat r}_{ij}\mathbf{\hat r}_{ij}}}\Big)  \notag\\ 
&{}+\frac{1}{2}\Big(\frac{a}{r_{ij}}\Big)^3\Big({\bf {1}}-3{{\mathbf{\hat r}_{ij}\mathbf{\hat r}_{ij}}}\Big)\bigg), 
\label{mu_tt} \\ 
\bm{\mu}_{ij}^{\mathrm{rr}}=&{}-\mu^\mathrm{r}\frac{1}{2}\left(\frac{a}{r_{ij}}\right)^3\left({\bf {1}}-3{{\mathbf{\hat r}_{ij}\mathbf{\hat r}_{ij}}}\right), \\ 
\bm{\mu}_{ij}^{\mathrm{tr}}=&\bm{\mu}_{ij}^{\mathrm{rt}}=\mu^\mathrm{r}\left(\frac{a}{r_{ij}}\right)^3{ {\mathbf r_{ij}}}\times, 
\label{mu_tr}
\end{align} 
with the distance vector $\vect{r}_{ij}=\vect{r}_j-\vect{r}_i$, $r_{ij}=|\vect{r}_{ij}|$ its absolute value,  and $\uvec{r}_{ij}=\vect{r}_{ij}/r_{ij} $.
The additional contributions due to the presence of the active force centers (derived from the previously introduced minimal microswimmer model) are given by \cite{menzel2016dynamical, hoell2017dynamical}
\begin{eqnarray}
\bm{\Lambda}_{ij}^{\mathrm{tt}} &=& \bm{\mu}_{ij}^{\mathrm{tt}+}-\bm{\mu}_{ij}^{\mathrm{tt}-},
\label{Lambda_tt} \\
\bm{\Lambda}_{ij}^{\mathrm{rt}} &=& \bm{\mu}_{ij}^{\mathrm{rt}+}-\bm{\mu}_{ij}^{\mathrm{rt}-}, 
\label{Lambda_rt}
\end{eqnarray}
with
\begin{eqnarray}
\bm{\mu}_{ij}^{\mathrm{tt}\pm} &=&
\frac{1}{8\pi\eta r_{ij}^{\pm}}\left({\bf {1}}+\mathbf{\hat r}_{ij}^{\pm}\mathbf{\hat r}_{ij}^{\pm}\right) \nonumber \\
&&{}+\frac{a^2}{24\pi\eta \left({r_{ij}^{\pm}}\right)^3}\left({\bf {1}}-3\mathbf{\hat r}^{\pm}_{ij}\mathbf{\hat r}^{\pm}_{ij}\right), 
\label{mu_tt_pm}\\
\bm{\mu}_{ij}^{\mathrm{rt}\pm} &=& \frac{1}{8\pi\eta \left({r_{ij}^{\pm}}\right)^3}\mathbf r_{ij}^{\pm}\times,
\label{mu_rt_pm}
\end{eqnarray} 
and 
\begin{eqnarray}
\mathbf r_{ij}^+ &=& \mathbf r_{ij}+\alpha L \mathbf{\hat n}_j, \label{defplus}
\\
\mathbf r_{ij}^- &=& \mathbf r_{ij}-(1-\alpha) L \mathbf{\hat n}_j. \label{defminus}
\end{eqnarray}
We neglect the distortion of the self-induced flow field that would result from the presence of the rigid spheres.
\cite{Kim_Karrila, adhyapak2017flow}$\kern -0.05em {}^{,}\kern -0.05em$\footnote{Strictly speaking, only the lowest-order terms in an expansion around $a/L=0$ are included for $j \neq i$ in Eqs.~(\ref{mu_tt_pm}) and (\ref{mu_rt_pm}).
We neglect higher-order corrections\cite{Kim_Karrila, adhyapak2017flow} in favor of simplicity and more traceable analytical expressions.}


\rev{Next, w}e specify the forces on the sphere representing the passive body of swimmer $j$ as
\begin{align}
 \vect{F}_j = &{}- \nabla_{\vect{r}_j} u_\mathrm{ext} (\vect{r}_j) 
	      - \nabla_{\vect{r}_j} \sum\limits_{k\neq j}  u(\vect{r}_j,\vect{r}_k) \notag \\ &{}- k_\mathrm{B} T \, \nabla_{\vect{r}_j} \ln P,
\end{align}
where $u_\mathrm{ext}(\vect{r})$ may include the effect of an external potential, $u(\vect{r}_j,\vect{r}_k)$ is a pairwise additive interaction potential, and the last term constitutes an entropic force that eventually leads to the correct diffusional parts of our statistical description. 
As usual, $k_\mathrm{B}$ denotes the Boltzmann constant and $T$ the temperature. 
The corresponding passive torques read
\begin{equation}
 \vect{T}_j = {}- k_\mathrm{B} T \, \uvec{n}_j \times \nabla_{\uvec{n}_j} \ln P, 
\end{equation}
consisting of only an entropic part, which likewise in the end correctly reproduces (rotational) diffusion.

To reduce the multi-dimensional nature of the probability density $P$ containing all $N$ swimmer coordinates $\vect{X}_i$, we intend to derive a dynamical equation only involving the reduced $n$-swimmer densities 
\begin{equation}
 \rh{n}(\vect{X}^n, t) = \frac{N!}{(N-n)!} \int \mathrm{d} \vect{X}_{n+1} ... \mathrm{d} \vect{X}_N P(\vect{X}^N,t).
\end{equation}
Particularly, we are interested in a dynamical equation for 
the one-swimmer density $\rh{1}(\vect{X},t)$.
As the swimmers are identical and, e.g., $\vect{X}$ in $\rh{1}(\vect{X},t)$ stands for the coordinate of ``one swimmer'' and not of ``swimmer $1$'', the enumeration $\vect{X}, \vect{X}', \vect{X}'', ...$ is used throughout this work when discussing arguments of $n$-swimmer densities.

\begin{widetext}
Integrating out the degrees of freedom $\vect{X}_i$ for all swimmers but one in Eq.~(\ref{Smoluchowski}), we obtain \cite{menzel2016dynamical, hoell2017dynamical}
\begin{equation}
\label{BBGKY1}
\frac{\partial\rho^{(1)}(\vect{X},t)}{\partial t} = {}-\nabla_{\vect{r}}\cdot(\curr{}^\mathrm{tt}+\curr{}^\mathrm{tr}+\curr{}^\mathrm{ta})
- \left(\uvec{n} \times \nabla_{\uvec{n}}\right) \cdot(\curr{}^\mathrm{rt}+\curr{}^\mathrm{rr}+\curr{}^\mathrm{ra}),
\end{equation}
with current densities \cite{menzel2016dynamical, hoell2017dynamical}
\begin{eqnarray}
\curr{}^\mathrm{tt}
&=&
{}-\mu^\mathrm{t}\left(k_\mathrm{B} T \nabla_{\mathbf r}\,\rh{1}(\vect{X},t)+\rh{1}(\vect{X},t) \nabla_{\mathbf r}\,u_\mathrm{ext}(\mathbf r) 
+\int \mathrm{d}\vect{X}' \rh{2}(\vect{X}, \vect{X}', t)\nabla_{\mathbf r}u(\mathbf r, \mathbf r')\right) 
\nonumber\\ &&
{}-\int \mathrm{d}\vect{X}' \,{\bm{\mu}^{\mathrm{tt}}_{\mathbf r, \mathbf r'}}\cdot\bigg(k_\mathrm{B} T \nabla_{\mathbf r'}\rh{2}(\vect{X},\vect{X}',t)
+\rh{2}(\vect{X},\vect{X}', t)\nabla_{\mathbf r'}u_\mathrm{ext}(\mathbf r')\nonumber\\ &&
\qquad{}+\rh{2}(\vect{X},\vect{X}', t)\nabla_{\mathbf r'}u({\mathbf r,\mathbf r')} 
+\int \mathrm{d}\vect{X}'' \rh{3}(\vect{X},\vect{X}',\vect{X}'', t)\nabla_{\mathbf r'}u(\mathbf r',\mathbf r'')\bigg),
\label{eqJ1} \\ 
\curr{}^\mathrm{tr}
&=&
{}-\int \mathrm{d}\vect{X}' \,k_\mathrm{B} T \bm{\mu}_{\mathbf r, \mathbf r'}^{\mathrm{tr}} (\mathbf{\hat n}'\times\nabla_{\mathbf{\hat n}'})\rh{2}(\vect{X},\vect{X}', t),
\label{eqJ2} \\ 
\curr{}^\mathrm{ta}
&=&
f\left({\bm{\Lambda}^{\mathrm{tt}}_{\mathbf r, \mathbf r}}\cdot\mathbf{\hat n} \rho^{(1)}(\vect{X}, t)   
+\int \mathrm{d}\vect{X}' \,{\bm{\Lambda}^{\mathrm{tt}}_{\mathbf r, \vect{X}'}}\cdot\mathbf{\hat n}'\rh{2}(\vect{X},\vect{X}', t)\right),
\label{eqJ3} \\ 
\curr{}^\mathrm{rt}
&=&
{}-\int \mathrm{d}\vect{X}' {\bm{\mu}^{\mathrm{rt}}_{\mathbf r, \mathbf r'}} \bigg(k_\mathrm{B} T \nabla_{\mathbf r'}\rho^{(2)}(\rev{\vect{X},\vect{X}'},t) 
+\rh{2}(\vect{X},\vect{X}', t)\nabla_{\mathbf r'}u_\mathrm{ext}(\mathbf r')
\nonumber \\ &&
\qquad{}+\rh{2}(\vect{X},\vect{X}', t)\nabla_{\mathbf r'}u({\mathbf r,\mathbf r')} + \int \mathrm{d}\vect{X}'' \rh{3}(\vect{X},\vect{X}',\vect{X}'', t)\nabla_{\mathbf r'}u(\mathbf r',\mathbf r'')\bigg),
\label{eqJ4} \\ 
\curr{}^\mathrm{rr}
&=&
{}-k_\mathrm{B} T \mu^\mathrm{r}\mathbf{\hat n}\times \nabla_{\mathbf{\hat n}} \rho^{(1)}(\vect{X}, t)-
\int \mathrm{d}\vect{X}' \,k_\mathrm{B} T \bm{\mu}^{\mathrm{rr}}_{\mathbf r, \mathbf r'}\cdot(\mathbf{\hat n}'\times\nabla_{\mathbf n'})\rh{2}(\vect{X},\vect{X}', t),
\label{eqJ5} \\ 
\curr{}^\mathrm{ra}
&=&
f\int \mathrm{d}\vect{X}' \,{\bm{\Lambda}^{\mathrm{rt}}_{\mathbf r, \vect{X}'}}\, \mathbf{\hat n}' \rh{2}(\vect{X},\vect{X}', t).
\label{eqJ6}
\end{eqnarray}
\end{widetext}

It is important to keep in mind that Eqs.~(\ref{BBGKY1})--(\ref{eqJ6}) form a non-closed set of equations, as the unknown higher-order densities $\rh{2}$ and $\rh{3}$ are needed as an input.
When a similar procedure is applied to Eq.~(\ref{Smoluchowski}) to find dynamical equations for, e.g., the two-swimmer density $\rh{2}$, next-higher orders appear, constituting an escalating loop typical for BBGKY-like hierarchies of equations. \cite{hansen1990theory} 
Therefore, a closure is needed by expressing the interaction terms in Eqs.~(\ref{eqJ1})--(\ref{eqJ6}) containing the two- and three-swimmer densities as functionals of only the one-swimmer density.
Dynamical density functional theory (DDFT) \cite{evans1979nature,evans1992density,marconi1999dynamic,marconi2000dynamic,archer2004dynamical,chan2005time,espanol2009derivation,evans2010density,lowen2010density,wittkowski2011dynamical} provides a well-established means for this purpose, where an approach for the present system was outlined in previous works. \cite{menzel2016dynamical, hoell2017dynamical}

Yet, our previous mean-field approach \cite{menzel2016dynamical,hoell2017dynamical} seems not to be sufficient to address the question below, namely, the question under which circumstances the swimmers develop collective polar orientational order.
Particularly, the interplay between the hydrodynamic interactions and the two-swimmer density in the equations above appears to be insufficiently resolved at the level of our previous mean-field- and Onsager-type formulation. 
Thus, a more refined version is needed, see below.


\rev{\section{Application to microswimmers confined to a plane}
\label{sec:application}
}

In the following, we consider microswimmers in suspension, yet with their positions $\vect{r}_i$ and orientations $\uvec{n}_i$, $i=1,\dots,N$,
confined to the flat $xy$-plane.
The surrounding fluid is still treated as three-dimensional.
Then, the orientation of each swimmer in Eqs.~(\ref{BBGKY1})--(\ref{eqJ6}) can be fully described by one angle $\phi_i$, 
and the orientational gradient operator becomes $\uvec{n}\times\nabla_\uvec{n}=\uvec{z} \partial_\phi$.
Such a system could be \rev{possibly} realized, e.g., by using optical trapping fields or by placing the swimmers at the interface between two immiscible fluids of identical viscosity.

Several further assumptions are introduced. 
First, the external potential shall vanish, i.e., $u_\mathrm{ext} = 0$.
Next, the system is confined to a two-dimensional box of area $A$ with periodic boundary conditions, containing our $N$ identical microswimmers.
We further assume that the one-swimmer density $\rh{1}(\vect{X},t)$, now with $\vect{X}=(\vect{r},\phi)$, is spatially homogeneous.\cite{lugo2016binary}
Thus, only variations as a function of the orientation variable $\phi$ are considered, i.e., $\rh{1}(\vect{X},t)=: A^{-1} \rh{1}(\phi,t)$, where the one-swimmer orientational density $\rh{1}(\phi,t)$ has been defined.

Eq.~(\ref{BBGKY1}) is now integrated over all spatial positions $\vect{r}$ in the area $A$.
Then the currents $\curr{}^\mathrm{tt}$, $\curr{}^\mathrm{tr}$, $\curr{}^\mathrm{ta}$ disappear from the equation
and the set of Eqs.~(\ref{BBGKY1})--(\ref{eqJ6}) is simplified to 
\begin{equation}
\label{BBGKY2}
\frac{\partial\rho^{(1)}(\phi,t)}{\partial t} = {}- \partial_\phi \int \mathrm{d} \vect{r} \, (\uvec{z}\cdot\curr{}^\mathrm{rt}+\uvec{z}\cdot\curr{}^\mathrm{rr}+\uvec{z}\cdot\curr{}^\mathrm{ra}).
\end{equation}
For spherical swimmer bodies, the integral term in Eq.~(\ref{eqJ5}) vanishes \cite{hoell2017dynamical} so that only the direct rotational diffusional part remains. 
Thus, Eq.~(\ref{BBGKY2}) can be rewritten as
\begin{widetext}
 \begin{equation}
\label{BBGKY3}
\frac{\partial\rho^{(1)}(\phi,t)}{\partial t} = D^\mathrm{r} \partial_\phi^2 \rh{1}(\phi,t)
 - f \partial_\phi \int \mathrm{d} \vect{r} \int \mathrm{d}\vect{X'} \, \uvec{z} \cdot \left({\vgr{\Lambda}^{\mathrm{rt}}_{\vect{r},\vect{X}'}}\,\uvec{n}'\right) \rho^{(2)}(\vect{X}, \vect{X}', t) 
- \partial_\phi \int \mathrm{d} \vect{r} \, \uvec{z}\cdot\curr{}^\mathrm{rt},
\end{equation}
where the last term approximately vanishes as detailed in App.~\ref{app:vanishingJ4} and $D^\mathrm{r}=k_\mathrm{B} T \mu^\mathrm{r}$ is the rotational diffusion constant for passive particles.

The remaining task is to find a reasonable approximation for $\rh{2}(\vect{X},\vect{X}',t)$.
Generally, the two-swimmer density is related to the one-swimmer density via 
$\rh{2}(\vect{X},\vect{X}',t)=\rh{1}(\vect{X},t) \, \rh{1}(\vect{X}',t) \, g^{(2)}(\vect{X},\vect{X}',t)$,
where $g^{(2)}(\vect{X},\vect{X}',t)$ is the pair distribution function.
Since we assume that the one-swimmer density does not depend on the spatial position,
this simplifies to  $\rh{2}(\vect{X},\vect{X}',t)=A^{-2} \rh{1}(\phi,t) \, \rh{1}(\phi',t) \, g^{(2)}(\vect{X},\vect{X}',t)$.
Furthermore, the pair distribution function in a spatially homogeneous system depends on only the \emph{relative} distance vector between the two particles,
so that $g^{(2)}(\vect{X},\vect{X}',t)=g^{(2)}(\vect{R},\phi,\phi',t)$ holds, with $\vect{R}:=\vect{r}'-\vect{r}$ the distance vector.
Thus, the second term on the right-hand side of Eq. (\ref{BBGKY3}) becomes
\begin{align}
I_1 := &{}-  f \partial_\phi \int \mathrm{d} \vect{r} \int \mathrm{d}\vect{r'} \int \mathrm{d}\phi' \, \uvec{z} \cdot \left({\vgr{\Lambda}^{\mathrm{rt}}_{\vect{r},\vect{X}'}}\,\uvec{n}'\right) \rho^{(2)}(\vect{X}, \vect{X}', t) \notag \\
= &{}- A^{-2} f \partial_\phi \left( \rh{1}(\phi,t)  \int \mathrm{d} \phi' \, \rh{1}(\phi',t) \int \mathrm{d} \vect{r} \int \mathrm{d}\vect{R}  \, \uvec{z} \cdot \left({\vgr{\Lambda}^{\mathrm{rt}}_{\vect{r},\vect{X}'}}\,\uvec{n}'\right) \, g^{(2)}(\vect{R},\phi,\phi',t) \right),
\label{def_I1}
\end{align}
where the spatial integral over $\vect{r}'$ has been shifted to $\vect{R}$.

To leading order in $R^{-1}$, with $R=|\vect{R}|$ the absolute value of the distance vector, the approximation
\begin{equation}
 \uvec{z} \cdot \left({\vgr{\Lambda}^{\mathrm{rt}}_{\vect{r},\vect{X}'}}\,\uvec{n}'\right) \approx -3 \mu^\mathrm{r} a^3L \cos(\phi'-\theta) \sin(\phi'-\theta)R^{-3}
\label{lambda_rt_expansion}
\end{equation}
holds, where $\theta$ is the angle between $\vect{R}$ and $\uvec{x}$, i.e., $\vect{R}=R (\cos \theta, \sin \theta)$. 
The orientation-dependent pair distribution function in the isotropic disordered state features a global rotational symmetry, i.e., it stays the same when we rotate the system by subtracting a common angle from all angles $\theta$, $\phi$, and $\phi'$. We select $\phi$ as that angle.
In other words, \rev{following standard arguments,} we may address the function in one particular frame of reference,\rev{\cite{Gray_Gubbins}} for which we now choose the frame of $\phi=0$.
In the following, $\gbody(R,\theta-\phi,\phi'-\phi)$ denotes the pair distribution function in this frame.
Moreover, the integral over $\vect{r}$ is now trivial, yielding the area $A$.
In combination, this leads to 
\begin{equation}
 I_1 \approx \frac{3 \mu^\mathrm{r} a^3L f} {A} \partial_\phi \left( \rh{1}(\phi,t)  \int \mathrm{d} \phi' \, \rh{1}(\phi',t) \int \mathrm{d}R \int \mathrm{d} \theta  \, \frac{\cos(\phi'-\theta) \sin(\phi'-\theta)}{R^{2}} \,  \gbody\left( R,\theta-\phi,\phi'-\phi,t \right) \right).
\label{I2_general}
\end{equation}

The starting point for all following considerations is thus the equation
 \begin{align}
\label{BBGKY4}
\frac{\partial\rho^{(1)}(\phi,t)}{\partial t} = &\, D^\mathrm{r} \partial_\phi^2 \rh{1}(\phi,t) +
 \frac{3 \mu^\mathrm{r} a^3L f} {A} \partial_\phi \bigg( \rh{1}(\phi,t)  \int \mathrm{d} \phi' \, \rh{1}(\phi',t) \notag \\ &\times \int \mathrm{d}R \int \mathrm{d} \theta \, \frac{\cos(\phi'-\theta) \sin(\phi'-\theta)}{R^{2}} \, \gbody\left( R,\theta-\phi,\phi'-\phi,t \right) \bigg) \notag \\
=: &\, D^\mathrm{r} \partial_\phi^2 \rh{1}(\phi,t) -
 3 \mu^\mathrm{r} a^3L f \frac{\rho_0}{N} \partial_\phi \left( \rh{1}(\phi,t)  \int \mathrm{d} \phi' \, \rh{1}(\phi',t) K(\phi-\phi',t) \right),
\end{align}
where we have introduced the global density $\rho_0=N / A$ and
further defined the function 
\begin{equation}
  K(\phi-\phi',t):={}-\int \mathrm{d}R \int \mathrm{d} \theta \, \frac{\cos(\phi'-\theta) \sin(\phi'-\theta)}{R^{2}} \, \gbody\left( R,\theta-\phi,\phi'-\phi,t \right),
\label{def_K}
\end{equation}
which represents a weighted integral of the pair distribution function over the distance vector.
If $\gbody\left( R,\theta-\phi,\phi'-\phi,t \right)$ is known,  $K(\phi-\phi',t)$  can be calculated.
In case this input is available, Eq.~(\ref{BBGKY4}) can serve as the starting point of a stability analysis of the isotropic disordered state, see Sec.~\ref{sec:linear} below.
\end{widetext}

From symmetry it follows that the simplest guess $g^{(2)}\equiv 1$ lets the second term on the right-hand side of Eq.~(\ref{BBGKY4}) vanish and is thus not sufficient to study the possible development of alignment.
As shown \rev{later}, an ansatz only featuring a spatial front--rear asymmetry, as previously used for a minimal mathematical description of motility-induced phase separation,\cite{bialke2013microscopic} also leads to a decay of any weak initial orientational order in a linear stability analysis of the isotropic disordered state. 
Thus, our next step is to address more carefully the pair distribution and to find approximate expressions in order to investigate the emergence of possible alignment.

\section{Approximation of the pair distribution function in the isotropic disordered state: DDFT and the Percus method}
\label{sec:pdf_approx}

\begin{figure}[th]
\vspace{-20pt}
\includegraphics{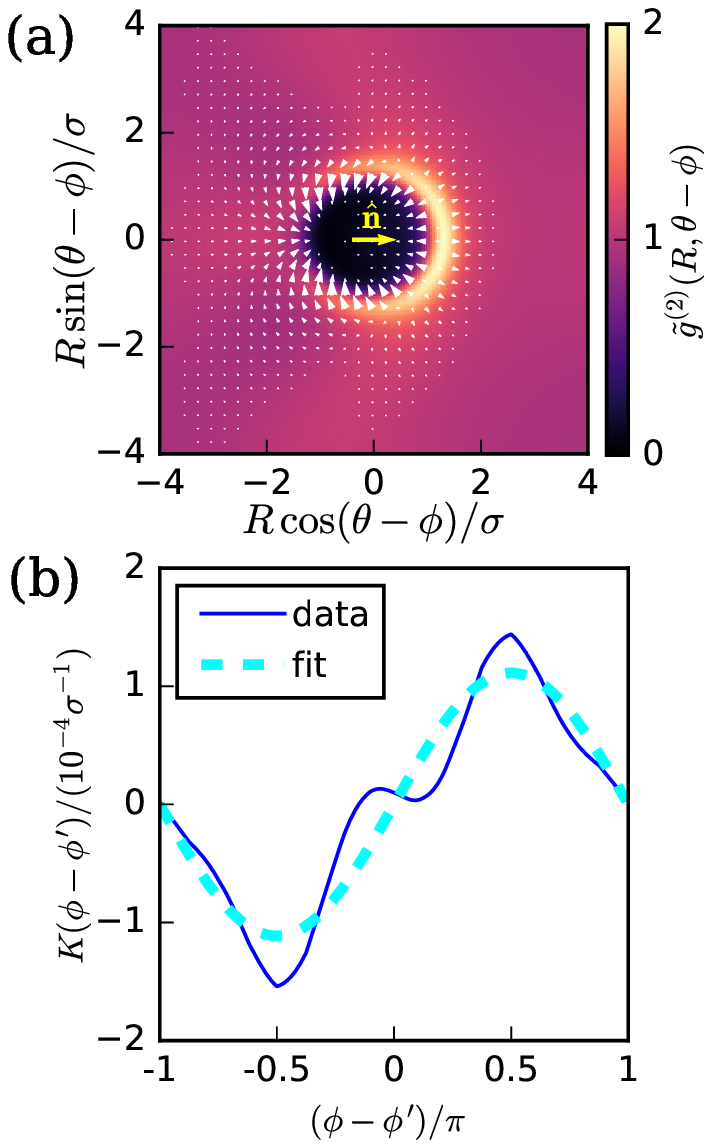}
\vspace{-10pt}
\caption{ 
(a) Swimmer--swimmer orientation-dependent pair distribution function, obtained via DDFT in combination with our adapted Percus test-particle method for active agents as described in the main text.
Brighter colors indicate a higher magnitude of the pair distribution function integrated over all orientations, i.e., we define \mbox{$\tilde{g}^{(2)}(R,\theta-\phi):=\int \mathrm{d} \phi' \gbody(R,\theta-\phi,\phi'-\phi)$}.
\rev{Thus, brighter colors imply} a higher probability to find a nearby swimmer.
White arrows mark the average orientations of nearby swimmers, calculated from $\int \mathrm{d} \phi' \uvec{n}'(\phi'-\phi) \gbody(R,\theta-\phi,\phi'-\phi)$. 
The large arrow at the center displays the orientation $\uvec{n}(\phi=0)$ of the fixed particle.
Parameter values are set to $\rho_0=0.0313 \sigma^{-2}$, $L=1.5\sigma$, $a=0.5\sigma$, $\alpha=0.4$, $V_0=20k_\mathrm{B}T$, and $f=50 k_\mathrm{B} T / \sigma$. 
The dimension of the square simulation box here is $8\sigma \times 8 \sigma$, and the DDFT equations are solved on a $128\times128\times16$ numerical grid for the discretization of $x$, $y$, and $\phi$ coordinates, respectively. 
Periodic boundary conditions were applied in all directions.
(b) Extracted function $K(\phi-\phi')$, defined in Eq.~(\ref{def_K}), for the same parameters as in (a).
Fitting with the function $C \sin(\phi-\phi')$ (dashed line) here leads to $C \approx 1.11 \times 10^{-4} \sigma^{-1}$.}
\label{fig:DDFT_percus}
\end{figure}

Our goal in this section is to identify a reasonable approximation for the pair distribution function of microswimmers in an isotropic disordered suspension to enable our subsequent study of the linear stability of the disordered state \rev{in Sec.~\ref{sec:linear}}. 
For this purpose, we \rev{here} adapt the Percus method,\cite{percus1962approximation} which is exact in equilibrium isotropic systems.
Yet, it should at least qualitatively hint
at the basic shape of the pair distribution in our inherently non-equilibrium system of self-propelled microswimmers.
Since a coarse knowledge of the general shape is sufficient for our objective, \rev{as well as for technical reasons detailed below}, hydrodynamic interactions are neglected \rev{throughout the present} section for simplicity.
\rev{That is,} approximations for the pair distribution function of ``dry'' self-propelled particles are determined. 
For strong force dipoles and in aligned systems, deviations from these reduced expressions \rev{will occur}.\cite{pessot2018}

\rev{
\subsection{The Percus method}
}
In the Percus method for fluids in equilibrium, \cite{percus1962approximation} one particle is declared a test particle and fixed in (phase) space, e.g., at position $\vect{r}$.
Then its effect on the remaining particles is effectively described as an external potential.
Percus showed that in a homogeneous fluid the resulting inhomogeneous density distribution of the other particles at positions $\vect{r}'$ around the first particle is connected to the pair distribution function via
the exact relation \mbox{$\rho(\vect{r}'-\vect{r})=\rho_0 g^{(2)}(\vect{r}'-\vect{r})$}, where $\rho_0$ is the (constant) overall density of the bulk fluid.
This way, the pair distribution function of a liquid \rev{equilibrium} system can be obtained.

A recent equilibrium classical density functional theory study shows that employing 
the Percus method can lead to good approximations of pair distribution functions, even if using a simple mean-field approximation for the excess functional. \cite{archer2017standard}
In the past, some studies have addressed dynamical test-particle methods for passive particles.\cite{hopkins2010van,brader2015power}
Nevertheless, it is still an open question how good of an approximation this method is for an active non-equilibrium system (as ours).
This should be examined in detail in future work and compared to other approaches. \cite{rein2016applicability, hartel2018three}

For \rev{a reasonable} description of the pair distribution function, we additionally need to account for the orientational degree(s) of freedom and the self-propulsion of the test particle.
The latter can be achieved by switching to the body frame of the test particle and ``streaming'' all other particles
oppositely to the (fixed) swimming direction of the test particle with its effective swimming speed $v_\mathrm{s}$.
In a non-dilute system, interactions between the ``non-test'' particles can be included via DDFT. \cite{evans1979nature,evans1992density,marconi1999dynamic,marconi2000dynamic,archer2004dynamical,chan2005time,espanol2009derivation,evans2010density,lowen2010density,wittkowski2011dynamical,menzel2016dynamical}
By definition, $0<v_\mathrm{s}\leq v_0$ holds in our ``dry'' system, with $v_0$ the free swimming speed of an unconstricted single swimmer.
In very dense cases of swimming being blocked by the presence of other particles, $v_\mathrm{s} \to 0$ is also possible (over a certain interval, $v_\mathrm{s}$ will decline approximately linearly with increasing local density\cite{stenhammar2013continuum, bialke2013microscopic}).  

We select the orientation of the fixed particle as $\phi=0$. 
The sign of $f$ then determines the angle $\psi$ of swimming given by $\psi=\phi$ for pushers and $\psi=\phi+\pi$ for pullers.
Thus, $\vect{v}_\mathrm{st}:={}- \mathrm{sign}(f) v_\mathrm{s} \uvec{x}$ is the additional velocity with which the other particles are streamed against the first, fixed particle.
Here, we choose $v_\mathrm{s}=v_0$, which is appropriate for dilute systems.

\rev{
\subsection{Evaluation using DDFT}
}

Now we follow our previous works \cite{menzel2016dynamical,hoell2017dynamical} \rev{for} (numerically) implementing the DDFT \rev{(neglecting hydrodynamic interactions as mentioned above)}.\cite{wensink2008aggregation}
Formally, this means that the tensors
 ${\bm{\mu}^{\mathrm{tt}}_{\mathbf r, \mathbf r'}}$,
$\bm{\mu}_{\mathbf r, \mathbf r'}^{\mathrm{tr}}$,
${\bm{\Lambda}^{\mathrm{tt}}_{\mathbf r, \vect{X}'}}$,
${\bm{\mu}^{\mathrm{rt}}_{\mathbf r, \mathbf r'}}$,
$\bm{\mu}^{\mathrm{rr}}_{\mathbf r, \mathbf r'}$,
and ${\bm{\Lambda}^{\mathrm{rt}}_{\mathbf r, \vect{X}'}}$ 
in Eqs.~(\ref{eqJ1})--(\ref{eqJ6})
are all set to zero.
Without hydrodynamic interactions, the only difference between pusher and puller microswimmers is that a corresponding swimmer propels into the direction of $\uvec{n}$ or, respectively, ${}-\uvec{n}$, see Fig.~\ref{fig:model}.
The steric interaction potential between swimmers $i$ and $j$ is now specified as the GEM-4 potential \cite{mladek2006formation,archer2014solidification} with
\begin{equation}
 u(\vect{r}_i,\vect{r}_j)=V_0 \exp\left(-\left(\frac{r_{ij}}{\sigma}\right)^4\right),
\label{GEM}
\end{equation}
where $V_0$ describes the strength of the potential.

Consequently, the potential $u(\vect{0},\vect{r})$ following from Eq.~(\ref{GEM}) is used as the external potential $u_\mathrm{ext}(\vect{r})$ in Eq.~(\ref{eqJ1}) when evaluating our DDFT.
It represents the fixed particle at the origin used in the Percus method. 
Furthermore, the streaming of all other swimmers, as described above, is enforced by applying an additional constant force $\nabla_\vect{r} u_\mathrm{ext}(\vect{r}) = {}-\vect{v}_\mathrm{st} / \mu^\mathrm{t}$ in Eq.~(\ref{eqJ1}), which continuously drives the particle density against the test particle and across the periodic boundaries.
At this point, it also becomes obvious why including the hydrodynamic interactions \rev{in this method would lead to challenging problems}. 
\rev{If hydrodynamic interactions were present}, simply including the streaming velocity $\vect{v}_\mathrm{st}$ as indicated above \rev{would} 
neglect the hydrodynamic interactions resulting from \rev{the flow fields that} the test swimmer and the other swimmers \rev{generate during their active motion}. 
\rev{Moreover}, driving swimmers towards each other \rev{by net forces to mimic their mutual approach during self-propulsion would induce unphysical} fluid flows. 
\rev{The hydrodynamic interactions} resulting from such net forces (force monopoles) are different from the actual ones \rev{resulting from} force dipoles. 
Clearly, this opens the way for additional studies in the future \rev{to address these issues}. 
At our present level of searching for the leading-order angular dependence \rev{of the pair distribution function}, neglecting the hydrodynamic interactions \rev{appears viable}, see below.

For consistency, the interaction strength $V_0$ must be sufficiently high to hinder other particles from swimming or being streamed through the fixed particle. 
Repeating the choice of our previous works, again the mean-field functional is employed to specify the corresponding excess free energy in the DDFT. 
Then, the DDFT equations are solved numerically using a finite-volume method solver \cite{Guyer_2009_CiSE} until a steady state is reached. 
This steady state describes the orientation-dependent particle distribution function (with $\phi=0$) that we searched for.

\rev{
\subsection{Resulting functional form}
}
Figure~\ref{fig:DDFT_percus}(a) shows a typical pair distribution function obtained in this way for non-hydrodynamically-interacting pushers in the isotropic disordered state. 
We find qualitative agreement with previous (orientationally-averaged) pair distribution functions of self-propelled agents determined by particle-based computer simulations,\cite{bialke2013microscopic,pessot2018}
e.g., concerning the front--rear asymmetry.
The extracted function $K(\phi-\phi')$ is displayed in Fig.~\ref{fig:DDFT_percus}(b).
For pullers of identical $|f|$, an analogous picture is found (as mentioned above, hydrodynamic interactions are not taken into account at the moment). In the end, an identical $K(\phi-\phi')$ is obtained.

Figure~\ref{fig:DDFT_percus}(b), determined in this way, demonstrates a dominant sinusoidal first-harmonic contribution in $K(\phi-\phi')$.
We thus to lowest order approximate
\begin{equation}
 K(\phi-\phi') \approx C \sin(\phi-\phi'),  \, \mathrm{with}  \, C>0.
\label{K_sinus}
\end{equation}
The amplitude $C$ has the dimension of inverse length and depends in a non-trivial way on $v_\mathrm{s}$, $\rho_0$, and the microscopic parameters in the swimmer model. 

Since the anisotropy of the pair distribution function is most pronounced near the surface of the fixed particle, see Fig.~\ref{fig:DDFT_percus}(a),
this angular dependence of $K(\phi-\phi')$ seems to be effectively caused by the short-range steric interaction.
Thus, point particles may not show the type of behavior identified in Sec.~\ref{sec:linear} below.\cite{stenhammar2017role}
A corresponding dominance of the steric interactions \rev{at least} supports neglecting the hydrodynamic interactions in the treatment above to lowest order.

\rev{Moreover, the functional form of $K(\phi-\phi')$ in Eq.~(\ref{K_sinus}) can also be motivated in a different way for
dilute systems as ours, see App.~\ref{app:scattering}.
Accordingly, our result above is supported by an independent approach.}
\rev{A further confirmation of the form in Eq.~(\ref{K_sinus}) is given in App.~\ref{app:analytic}.}

\section{Linear stability analysis}
\label{sec:linear}

Finally, we \rev{now test} for the linear instability of the isotropic disordered microswimmer system.
\rev{For this purpose, we turn back to Eqs.~(\ref{BBGKY1})--(\ref{BBGKY4}) that explicitly include hydrodynamic interactions via the hydrodynamic mobility tensors.}
\rev{Nevertheless, in the absence of a more sophisticated approximation, we assume the functional form in Eq.~(\ref{K_sinus}) found for neglected hydrodynamic interactions and use it as an input to these equations to} check whether collective orientational order spontaneously arises from a linear instability of the state of absent orientational order.

As further elucidated in App.~\ref{appendix_rho0}, the static uniform distribution $\rho(\phi,t) = N (2 \pi)^{-1}$~is always a solution of Eq.~(\ref{BBGKY4}).
However, \rev{as shown in the following,} it is either linearly stable or unstable, \rev{depending on the system parameters}.
If it is linearly stable, the system remains in the isotropic disordered state for that set of parameter values, at least in the absence of larger fluctuations, perturbations, and spatial inhomogeneities.
If it is linearly unstable, it will develop a different state, e.g., one of collective polar order. 
To test for linear stability, a small harmonic fluctuation is superimposed onto the uniform distribution, 
i.e., 
$\rho(\phi,t)=N (2 \pi)^{-1} + \epsilon(t) \cos(\phi-\phi_0)$, 
with small $\epsilon(t) \ll N (2 \pi)^{-1}$ and arbitrary $\phi_0$.

This ansatz is inserted into Eq.~(\ref{BBGKY4}).
Via Eq.~(\ref{K_sinus}), two terms vanish due to symmetry upon performing the integration, one term can be neglected via $\epsilon^2 (t) \ll \epsilon (t)$, and we arrive at
\begin{equation}
 \dot{\epsilon}(t) \cos(\phi-\phi_0) = {}- D^\mathrm{r} \epsilon(t) \cos(\phi-\phi_0) + \tilde{I}_1 \epsilon(t)
\label{stability}
\end{equation}
with a dot denoting a time derivative and
\begin{equation}
 \tilde{I}_1 := {}- \frac{3 \mu^\mathrm{r} a^3L f \rho_0}{2 \pi} \;  \partial_\phi \left(   \int \mathrm{d} \phi' \, \cos(\phi'-\phi_0) K(\phi - \phi') \right).
\end{equation}
Using Eq.~(\ref{K_sinus}), 
this simplifies to 
\begin{align}
 \tilde{I}_1 	= &{}- \frac{3}{2} \mu^\mathrm{r} a^3L C f \rho_0 \cos(\phi-\phi_0).
\label{tildeI2}
\end{align}
Combining Eqs.~(\ref{stability}) and (\ref{tildeI2}) leads to the ordinary differential equation
\begin{equation}
  \dot{\epsilon}(t) = \left(- D^\mathrm{r} - \frac{3}{2} \mu^\mathrm{r} a^3L C \rho_0 f \right) \epsilon(t).
\label{linear_ode}
\end{equation}
Its solution for the amplitude $\epsilon(t)$ of the perturbation is an exponential function that decays in time when the bracketed term is negative, and grows otherwise.
For pushers ($f>0$), the fluctuation thus always decays ($\mu^\mathrm{r}, a, L, C, \rho_0$ are all positive).
In contrast to that, strong pullers with 
\begin{equation}
 f L< {}- \frac{2}{3} \frac{D^\mathrm{r}}{\mu^\mathrm{r} a^3 \rho_0  C} = {}- \frac{2}{3} \frac{k_\mathrm{B} T}{ a^3 \rho_0  C}
\label{growth_criterion}
\end{equation}
show exponential growth of fluctuations involving polar orientational order, i.e., the isotropic disordered state is linearly unstable against initial polar ordering.

We remark that, while an increased density $\rho_0$ in Eq.~(\ref{growth_criterion}) seems to support the emergence of orientational order, it is to be noted that $C$ heavily depends on the system parameters, including $\rho_0$, and can overshadow that effect. 
For instance, at high densities, the swimmers may mutually disturb and block their motion. 
Then, the global orientational dependence of the pair distribution function should change, possibly implying $C \to 0$. 
This would counteract the emergence of a global polar ordering via the mechanism described in this work.
However, spatial variations would then certainly become important and should be included into the theoretical consideration as a possible future extension.

\section{Conclusions}
\label{sec:conclusions}

In summary, we have presented a microscopic statistical approach to describing and predicting the emergence of collective polar ordering in (semi-)dilute suspensions of \rev{active} force-dipole microswimmers in suspension. 
We found that such a polar order can arise in systems of pullers of strong enough activity to overcome thermal dealignment caused by rotational diffusion. 
Our statistical approach traces back the self-ordering of the system to the actively induced hydrodynamic rotation--translation coupling between the swimmers.
To find a reasonable approximation for the involved pair distribution function, a technique combining DDFT and the Percus method (pinning one swimmer and treating it as an obstacle for the other swimmers) for an active system has been proposed, as well as intuitive arguments of broken symmetry.
As the central result, disordered suspensions of pushers in our approach were always found to be linearly stable against initial development of collective polar orientational order.
In contrast to that, suspensions of strong pullers were observed to be linearly unstable against polar orientational ordering.
It will be interesting to further challenge our adapted test-particle method by quantitative comparison with simulations or other theoretical methods\cite{bialke2013microscopic,rein2016applicability, hartel2018three,nadine2018static,pessot2018} in the future.
\rev{Additionally, it would be intriguing to test the applicability of our approach and results as input for further studies on the mesoscale hydrodynamic behavior of microswimmer suspensions, possibly even concerning mesoscale turbulence.\cite{heidenreich2016hydrodynamic,reinken2018derivation}}

We \rev{wish} to remark that our system \rev{when} taken to the thermodynamic limit ($N\to \infty$ \rev{and} $A \to \infty$, while the average density is kept constant) \rev{might} still develop overall orientational order, against the Mermin-Wagner theorem. \cite{mermin1966absence} 
This is because of its inherently non-equilibrium nature. \cite{toner1995long,toner2005hydrodynamics} 
Nevertheless, additional spatially-resolved investigations would be very interesting as they could be able to discern between local and global ordering and show their interplay.

Furthermore, the theory can also be generalized to binary mixtures of different swimmer species, resulting in two coupled equations similar to Eq.~(\ref{BBGKY4}). Each of them contains an additional coupling term including the one-swimmer density of the other swimmer species. The results could then be compared with previous particle-based computer simulations of binary pusher--puller mixtures. \cite{pessot2018}
Apart from that, an extension to systems of hydrodynamically interacting self-propelled rods \cite{saintillan2007orientational} is conceivable as well.

\acknowledgments

The authors thank Giorgio Pessot for helpful discussions.
Support of this work by the Deutsche Forschungsgemeinschaft through the priority program SPP 1726 on microswimmers, grant nos.\ LO 418/17-2 and ME 3571/2-2, is gratefully acknowledged.

\appendix

\begin{widetext}
\section{}
\label{app:vanishingJ4}

In this appendix, we briefly demonstrate that the last term in Eq.~(\ref{BBGKY3}) vanishes approximately.
Regarding the current density \rev{$\curr{}^\mathrm{rt}$} defined in Eq.~(\ref{eqJ4}), the second contribution drops out because we here set $u_\mathrm{ext}=0$.
The third contribution vanishes for all isotropic central-force interaction potentials $u(\vect{r},\vect{r}')=u(|\vect{r}'-\vect{r}|)$
because the gradient of such a potential is parallel to the distance vector.
However, $\vgr{\mu}^\mathrm{rt}_{\vect{r},\vect{r}'}$ in Eq.~(\ref{eqJ4}) introduces the vector product with this distance vector, see Eq.~(\ref{mu_tr}), which then vanishes.
Finally, the contribution containing $\rh{3}(\vect{X},\vect{X}',\vect{X}'',t)$ in Eq.~(\ref{eqJ4}) is neglected for sufficiently dilute systems as it scales with  a higher order in $\rho_0$ than the other contributions.
Together, this reduces the last term of Eq.~(\ref{BBGKY3}) to
\begin{equation}
I_2 := - \partial_\phi \int \mathrm{d} \vect{r} \, \uvec{z}\cdot\curr{}^\mathrm{rt} \approx 
k_\mathrm{B} T \partial_\phi \int \mathrm{d} \vect{r} \int \mathrm{d}\vect{X}' \, \uvec{z} \cdot \bigg({\bm{\mu}^\mathrm{rt}_{\mathbf r, \mathbf r'}} \nabla_{\mathbf r'}\rho^{(2)}(\vect{X}, \vect{X}', t) \bigg),
\label{A1}
\end{equation}
which vanishes as is shown in the following. 

Using $\rh{2}(\vect{X},\vect{X}',t)=\rh{1}(\vect{X},t) \, \rh{1}(\vect{X}',t) \, g^{(2)}(\vect{X},\vect{X}',t)$ and $\rh{1}(\vect{X},t)= A^{-1} \rh{1}(\phi,t)$ as before,
Eq.~(\ref{A1}) can be rewritten as 
\begin{equation}
      I_2 \approx \frac{k_\mathrm{B} T}{A^2}  \partial_\phi \Bigg(\rh{1}(\phi,t) \int \mathrm{d} \vect{r} \int \mathrm{d} \phi' \rh{1}(\phi',t) \int \mathrm{d} \vect{r}' \mu^\mathrm{r} a^3 |\vect{r}'-\vect{r}|^{-3} \, \uvec{z} \cdot \left( (\vect{r}'-\vect{r})  \times \nabla_{\vect{r}'-\vect{r}} \, g^{(2)}(\vect{r}'-\vect{r},\phi,\phi',t) \right) \Bigg).
\end{equation}
The inner spatial integral is then transformed to the polar coordinates $(R,\theta)$, with 
$\vect{R}=\vect{r}'-\vect{r}=:R ( \cos\theta, \sin\theta )$, 
yielding
\begin{equation}
      I_2 \approx \frac{D^\mathrm{r} a^3}{A^2}  \partial_\phi \Bigg(\rh{1}(\phi,t) \int \mathrm{d} \vect{r} \int \mathrm{d} \phi' \rh{1}(\phi',t) \int \mathrm{d} R \, R^{-2} \int \mathrm{d} \theta  \, \uvec{z} \cdot \left( \vect{R} \times \nabla_{\vect{R}} \, g^{(2)}(R,\theta,\phi,\phi',t) \right) \Bigg).
\end{equation}
Via the relation
$\uvec{z} \cdot \left( \vect{R} \times \nabla_{\vect{R}} \right) = \partial_\theta$ and the inherent periodicity of the pair distribution function with respect to the angular variables, the integral over $\theta$ leads to $I_2\approx 0$.
\end{widetext}

\section{Weak scattering}
\label{app:scattering}


\rev{Equation~(\ref{K_sinus}) can further be motivated for dilute systems as ours via a ``weak scattering'' approach, effectively including hydrodynamic interactions to an approximate extent.}
\rev{Here, we s}uppose that two microswimmers are located at arbitrary phase space positions $\vect{X}$ and $\vect{X}'$. 
We disregard all diffusional processes and any disturbing hydrodynamic interactions for almost all times so that the swimmers move along straight paths, with initial orientations $\uvec{n}$ and  $\uvec{n}'$.
In effect, their hydrodynamic interactions are considered to occur only once in time, at the moment when they come closest to each other.
Furthermore, we use the leading-order expansion of $\uvec{z} \cdot \left({\vgr{\Lambda}^{\mathrm{rt}}_{\vect{r},\vect{X}'}}\,\uvec{n}'\right)$ as given in Eq.~(\ref{lambda_rt_expansion}).
Then, the effective angular shift of the first swimmer due to the mutual hydrodynamic interaction between the swimmers is approximated as 
\begin{equation}
 \delta\phi:= -3 \mu^\mathrm{r} a^3 L f |\vect{R}_0|^{-3} \delta{t} \cos(\phi'-\theta_0) \sin(\phi'-\theta_0),
\label{scatter1}
\end{equation}
with a typical interaction time $\delta{t}$ assumed to be the same for all configurations. Additionally, $\vect{R}_0$ is the closest distance vector, with $\vect{R}_0=:|\vect{R}_0|(\cos\theta_0,\sin\theta_0)$.

For this vector, $\vect{R}_0 \cdot (\uvec{n}'-\uvec{n}) = 0 $ applies,
which leads to $\theta_0=(\phi+\phi')/2$.
Inserting this relation into Eq.~(\ref{scatter1}) leads to
\begin{equation}
 \delta\phi= \frac{3}{2} \mu^\mathrm{r} a^3 L f \delta{t} |\vect{R}_0|^{-3} \sin(\phi-\phi'),
\label{scatter2}
\end{equation}
which again implies mutual dealignment for pushers ($f>0$) and mutual alignment for pullers ($f<0$).
We remark that Eq.~(\ref{scatter2}) is compatible with Eq.~(\ref{K_sinus}), i.e., with $K(\phi-\phi') \approx C \sin(\phi-\phi')$, $C>0$.

\section{} 
\label{app:analytic}

\rev{In this appendix}, we consider some more aspects concerning the angular dependence of the pair distribution function $\gbody(R,\theta-\phi,\phi'-\phi)$ in the regarded isotropic disordered state, leading to Eq.~(\ref{K_sinus}).
From Eq.~(\ref{def_K}) it is obvious that homogeneous terms in $\gbody(R,\theta-\phi,\phi'-\phi)$ do not contribute to $K(\phi-\phi')$.
Moreover, since the hydrodynamic interactions decrease with increasing swimmer--swimmer distance, 
attention is now focused on the high-density ring of \rev{a radius} approximately \rev{equal to} the effective particle diameter $\sigma$, see Fig.~\ref{fig:DDFT_percus}(a). 

The pair distribution function shown in Fig.~\ref{fig:DDFT_percus}(a) features a front--rear asymmetry in the spatial distribution,
which can be phenomenologically addressed to lowest order by a term $\sim \cos(\theta-\psi)$, where $\psi$ denotes the angle of the swimming direction as before.
Furthermore, the orientational distribution of nearby swimmers around the central swimmer seems to point inward, see the innermost white arrows in Fig.~\ref{fig:DDFT_percus}(a). 
An orientational distribution peaked at $\psi'=\theta+\pi$ would reflect this and can be modeled by a contribution $\sim {}- \cos(\psi'-\theta)$.
Eventually, we notice that in the high-density area at the front of the central swimmer in Fig.~\ref{fig:DDFT_percus}(a), the surrounding swimmers are preferably oriented in the propulsion direction of the central swimmer. 
This can be represented by a term $\sim \cos(\theta-\psi) \cos(\psi'-\theta)$.
At the rear of the central swimmer, this term still maintains the preferred inward orientation of the surrounding swimmers in Fig.~\ref{fig:DDFT_percus}(a).

Taking into account the different terms described above, 
we investigate the ansatz \rev{
\begin{align}
 \gbody&(R,\theta-\phi,\phi'-\phi)\approx 1 + \delta(R-\sigma) \notag \\ &\times  \bigg(c_1 + c_2 \cos(\theta-\psi) - c_3 \cos(\psi'-\theta)   
 \notag \\ &\quad{}+ c_4 \cos(\theta-\psi) \cos(\psi'-\theta) \bigg),
\label{analytical_ansatz}
\end{align}
with $c_1,c_2,c_3,c_4>0$.}
Inserting it into Eq.~(\ref{def_K}), only the contribution $\sim c_4$ does not vanish,
but indeed is in agreement with Eq.~(\ref{K_sinus}) for $K(\phi-\phi')$.

\begin{widetext}
\section{} %
\label{appendix_rho0}

We here argue that the uniform distribution $\rh{1}(\phi,t)=N /(2\pi)$ is indeed an exact solution of Eq.~(\ref{BBGKY4}).
For $f=0$, the equilibrium case of passive spherical particles is recovered.
It is readily seen that in this case $\rh{1}(\phi,t)=N /(2\pi)$ solves Eq.~(\ref{BBGKY4}).
Otherwise, for $f \neq 0$, the only remaining term in Eq.~(\ref{BBGKY4}) is the activity-induced one stemming from $\curr{}^\mathrm{ra}$ in Eq.~(\ref{eqJ6}).

Evaluating this term in Eq.~(\ref{BBGKY4}) for $\rh{1}(\phi,t)=\rh{1}(\phi',t)=N (2\pi)^{-1}$ and disregarding all constants reduces our task to show that
\begin{equation}
 W:=\partial_\phi \bigg(  \int \mathrm{d} \phi' \, \int \mathrm{d}R \int \mathrm{d} \theta \, \frac{\cos(\phi'-\theta) \sin(\phi'-\theta)}{R^{2}} \, \gbody\left( R,\theta-\phi,\phi'-\phi \right) \bigg)
\end{equation}
vanishes. 
If the integrals over the angles $\phi'$ and $\theta$ are now shifted to the angles $\phi'-\phi$ and $\theta-\phi$, respectively, no formal dependence on $\phi$ remains after integration. 
Thus, $W$ indeed vanishes.
We remark that this result still holds when taking into account all orders in $R^{-1}$, e.g., starting from Eq.~(\ref{def_I1}).
\end{widetext}

\bibliography{dilute_alignment}

\begin{thebibliography}{117}%
\makeatletter
\providecommand \@ifxundefined [1]{%
 \@ifx{#1\undefined}
}%
\providecommand \@ifnum [1]{%
 \ifnum #1\expandafter \@firstoftwo
 \else \expandafter \@secondoftwo
 \fi
}%
\providecommand \@ifx [1]{%
 \ifx #1\expandafter \@firstoftwo
 \else \expandafter \@secondoftwo
 \fi
}%
\providecommand \natexlab [1]{#1}%
\providecommand \enquote  [1]{``#1''}%
\providecommand \bibnamefont  [1]{#1}%
\providecommand \bibfnamefont [1]{#1}%
\providecommand \citenamefont [1]{#1}%
\providecommand \href@noop [0]{\@secondoftwo}%
\providecommand \href [0]{\begingroup \@sanitize@url \@href}%
\providecommand \@href[1]{\@@startlink{#1}\@@href}%
\providecommand \@@href[1]{\endgroup#1\@@endlink}%
\providecommand \@sanitize@url [0]{\catcode `\\12\catcode `\$12\catcode
  `\&12\catcode `\#12\catcode `\^12\catcode `\_12\catcode `\%12\relax}%
\providecommand \@@startlink[1]{}%
\providecommand \@@endlink[0]{}%
\providecommand \url  [0]{\begingroup\@sanitize@url \@url }%
\providecommand \@url [1]{\endgroup\@href {#1}{\urlprefix }}%
\providecommand \urlprefix  [0]{URL }%
\providecommand \Eprint [0]{\href }%
\providecommand \doibase [0]{http://dx.doi.org/}%
\providecommand \selectlanguage [0]{\@gobble}%
\providecommand \bibinfo  [0]{\@secondoftwo}%
\providecommand \bibfield  [0]{\@secondoftwo}%
\providecommand \translation [1]{[#1]}%
\providecommand \BibitemOpen [0]{}%
\providecommand \bibitemStop [0]{}%
\providecommand \bibitemNoStop [0]{.\EOS\space}%
\providecommand \EOS [0]{\spacefactor3000\relax}%
\providecommand \BibitemShut  [1]{\csname bibitem#1\endcsname}%
\let\auto@bib@innerbib\@empty
\bibitem [{\citenamefont {Purcell}(1977)}]{purcell1977life}%
  \BibitemOpen
  \bibfield  {author} {\bibinfo {author} {\bibfnamefont {E.~M.}\ \bibnamefont
  {Purcell}},\ }\href@noop {} {\bibfield  {journal} {\bibinfo  {journal} {Am.
  J. Phys.}\ }\textbf {\bibinfo {volume} {45}},\ \bibinfo {pages} {3} (\bibinfo
  {year} {1977})}\BibitemShut {NoStop}%
\bibitem [{\citenamefont {Howse}\ \emph {et~al.}(2007)\citenamefont {Howse},
  \citenamefont {Jones}, \citenamefont {Ryan}, \citenamefont {Gough},
  \citenamefont {Vafabakhsh},\ and\ \citenamefont
  {Golestanian}}]{howse2007self}%
  \BibitemOpen
  \bibfield  {author} {\bibinfo {author} {\bibfnamefont {J.~R.}\ \bibnamefont
  {Howse}}, \bibinfo {author} {\bibfnamefont {R.~A.}\ \bibnamefont {Jones}},
  \bibinfo {author} {\bibfnamefont {A.~J.}\ \bibnamefont {Ryan}}, \bibinfo
  {author} {\bibfnamefont {T.}~\bibnamefont {Gough}}, \bibinfo {author}
  {\bibfnamefont {R.}~\bibnamefont {Vafabakhsh}}, \ and\ \bibinfo {author}
  {\bibfnamefont {R.}~\bibnamefont {Golestanian}},\ }\href@noop {} {\bibfield
  {journal} {\bibinfo  {journal} {Phys. Rev. Lett.}\ }\textbf {\bibinfo
  {volume} {99}},\ \bibinfo {pages} {048102} (\bibinfo {year}
  {2007})}\BibitemShut {NoStop}%
\bibitem [{\citenamefont {Lauga}\ and\ \citenamefont
  {Powers}(2009)}]{lauga2009hydrodynamics}%
  \BibitemOpen
  \bibfield  {author} {\bibinfo {author} {\bibfnamefont {E.}~\bibnamefont
  {Lauga}}\ and\ \bibinfo {author} {\bibfnamefont {T.~R.}\ \bibnamefont
  {Powers}},\ }\href@noop {} {\bibfield  {journal} {\bibinfo  {journal} {Rep.
  Prog. Phys.}\ }\textbf {\bibinfo {volume} {72}},\ \bibinfo {pages} {096601}
  (\bibinfo {year} {2009})}\BibitemShut {NoStop}%
\bibitem [{\citenamefont {Elgeti}, \citenamefont {Winkler},\ and\ \citenamefont
  {Gompper}(2015)}]{elgeti2015physics}%
  \BibitemOpen
  \bibfield  {author} {\bibinfo {author} {\bibfnamefont {J.}~\bibnamefont
  {Elgeti}}, \bibinfo {author} {\bibfnamefont {R.~G.}\ \bibnamefont {Winkler}},
  \ and\ \bibinfo {author} {\bibfnamefont {G.}~\bibnamefont {Gompper}},\
  }\href@noop {} {\bibfield  {journal} {\bibinfo  {journal} {Rep. Prog. Phys.}\
  }\textbf {\bibinfo {volume} {78}},\ \bibinfo {pages} {056601} (\bibinfo
  {year} {2015})}\BibitemShut {NoStop}%
\bibitem [{\citenamefont {Z{\"o}ttl}\ and\ \citenamefont
  {Stark}(2016)}]{zottl2016emergent}%
  \BibitemOpen
  \bibfield  {author} {\bibinfo {author} {\bibfnamefont {A.}~\bibnamefont
  {Z{\"o}ttl}}\ and\ \bibinfo {author} {\bibfnamefont {H.}~\bibnamefont
  {Stark}},\ }\href@noop {} {\bibfield  {journal} {\bibinfo  {journal} {J.
  Phys.: Condens. Matter}\ }\textbf {\bibinfo {volume} {28}},\ \bibinfo {pages}
  {253001} (\bibinfo {year} {2016})}\BibitemShut {NoStop}%
\bibitem [{\citenamefont {Bechinger}\ \emph {et~al.}(2016)\citenamefont
  {Bechinger}, \citenamefont {Di~Leonardo}, \citenamefont {L{\"o}wen},
  \citenamefont {Reichhardt}, \citenamefont {Volpe},\ and\ \citenamefont
  {Volpe}}]{bechinger2016active}%
  \BibitemOpen
  \bibfield  {author} {\bibinfo {author} {\bibfnamefont {C.}~\bibnamefont
  {Bechinger}}, \bibinfo {author} {\bibfnamefont {R.}~\bibnamefont
  {Di~Leonardo}}, \bibinfo {author} {\bibfnamefont {H.}~\bibnamefont
  {L{\"o}wen}}, \bibinfo {author} {\bibfnamefont {C.}~\bibnamefont
  {Reichhardt}}, \bibinfo {author} {\bibfnamefont {G.}~\bibnamefont {Volpe}}, \
  and\ \bibinfo {author} {\bibfnamefont {G.}~\bibnamefont {Volpe}},\
  }\href@noop {} {\bibfield  {journal} {\bibinfo  {journal} {Rev. Mod. Phys.}\
  }\textbf {\bibinfo {volume} {88}},\ \bibinfo {pages} {045006} (\bibinfo
  {year} {2016})}\BibitemShut {NoStop}%
\bibitem [{\citenamefont {Eisenbach}\ and\ \citenamefont
  {Giojalas}(2006)}]{eisenbach2006sperm}%
  \BibitemOpen
  \bibfield  {author} {\bibinfo {author} {\bibfnamefont {M.}~\bibnamefont
  {Eisenbach}}\ and\ \bibinfo {author} {\bibfnamefont {L.~C.}\ \bibnamefont
  {Giojalas}},\ }\href@noop {} {\bibfield  {journal} {\bibinfo  {journal} {Nat.
  Rev. Mol. Cell Biol.}\ }\textbf {\bibinfo {volume} {7}},\ \bibinfo {pages}
  {276} (\bibinfo {year} {2006})}\BibitemShut {NoStop}%
\bibitem [{\citenamefont {Berg}(2008)}]{berg2008coli}%
  \BibitemOpen
  \bibfield  {author} {\bibinfo {author} {\bibfnamefont {H.~C.}\ \bibnamefont
  {Berg}},\ }\href@noop {} {\emph {\bibinfo {title} {E.\ coli in Motion}}}\
  (\bibinfo  {publisher} {Springer Science \& Business Media, New York},\
  \bibinfo {year} {2008})\BibitemShut {NoStop}%
\bibitem [{\citenamefont {Polin}\ \emph {et~al.}(2009)\citenamefont {Polin},
  \citenamefont {Tuval}, \citenamefont {Drescher}, \citenamefont {Gollub},\
  and\ \citenamefont {Goldstein}}]{polin2009chlamydomonas}%
  \BibitemOpen
  \bibfield  {author} {\bibinfo {author} {\bibfnamefont {M.}~\bibnamefont
  {Polin}}, \bibinfo {author} {\bibfnamefont {I.}~\bibnamefont {Tuval}},
  \bibinfo {author} {\bibfnamefont {K.}~\bibnamefont {Drescher}}, \bibinfo
  {author} {\bibfnamefont {J.~P.}\ \bibnamefont {Gollub}}, \ and\ \bibinfo
  {author} {\bibfnamefont {R.~E.}\ \bibnamefont {Goldstein}},\ }\href@noop {}
  {\bibfield  {journal} {\bibinfo  {journal} {Science}\ }\textbf {\bibinfo
  {volume} {325}},\ \bibinfo {pages} {487} (\bibinfo {year}
  {2009})}\BibitemShut {NoStop}%
\bibitem [{\citenamefont {Mussler}\ \emph {et~al.}(2013)\citenamefont
  {Mussler}, \citenamefont {Rafa{\"\i}}, \citenamefont {Peyla},\ and\
  \citenamefont {Wagner}}]{mussler2013effective}%
  \BibitemOpen
  \bibfield  {author} {\bibinfo {author} {\bibfnamefont {M.}~\bibnamefont
  {Mussler}}, \bibinfo {author} {\bibfnamefont {S.}~\bibnamefont {Rafa{\"\i}}},
  \bibinfo {author} {\bibfnamefont {P.}~\bibnamefont {Peyla}}, \ and\ \bibinfo
  {author} {\bibfnamefont {C.}~\bibnamefont {Wagner}},\ }\href@noop {}
  {\bibfield  {journal} {\bibinfo  {journal} {EPL (Europhys. Lett.)}\ }\textbf
  {\bibinfo {volume} {101}},\ \bibinfo {pages} {54004} (\bibinfo {year}
  {2013})}\BibitemShut {NoStop}%
\bibitem [{\citenamefont {Goldstein}(2015)}]{goldstein2015green}%
  \BibitemOpen
  \bibfield  {author} {\bibinfo {author} {\bibfnamefont {R.~E.}\ \bibnamefont
  {Goldstein}},\ }\href@noop {} {\bibfield  {journal} {\bibinfo  {journal}
  {Annu. Rev. Fluid Mech.}\ }\textbf {\bibinfo {volume} {47}} (\bibinfo {year}
  {2015})}\BibitemShut {NoStop}%
\bibitem [{\citenamefont {Paxton}\ \emph {et~al.}(2004)\citenamefont {Paxton},
  \citenamefont {Kistler}, \citenamefont {Olmeda}, \citenamefont {Sen},
  \citenamefont {St.~Angelo}, \citenamefont {Cao}, \citenamefont {Mallouk},
  \citenamefont {Lammert},\ and\ \citenamefont {Crespi}}]{paxton2004catalytic}%
  \BibitemOpen
  \bibfield  {author} {\bibinfo {author} {\bibfnamefont {W.~F.}\ \bibnamefont
  {Paxton}}, \bibinfo {author} {\bibfnamefont {K.~C.}\ \bibnamefont {Kistler}},
  \bibinfo {author} {\bibfnamefont {C.~C.}\ \bibnamefont {Olmeda}}, \bibinfo
  {author} {\bibfnamefont {A.}~\bibnamefont {Sen}}, \bibinfo {author}
  {\bibfnamefont {S.~K.}\ \bibnamefont {St.~Angelo}}, \bibinfo {author}
  {\bibfnamefont {Y.}~\bibnamefont {Cao}}, \bibinfo {author} {\bibfnamefont
  {T.~E.}\ \bibnamefont {Mallouk}}, \bibinfo {author} {\bibfnamefont {P.~E.}\
  \bibnamefont {Lammert}}, \ and\ \bibinfo {author} {\bibfnamefont {V.~H.}\
  \bibnamefont {Crespi}},\ }\href@noop {} {\bibfield  {journal} {\bibinfo
  {journal} {J. Am. Chem. Soc.}\ }\textbf {\bibinfo {volume} {126}},\ \bibinfo
  {pages} {13424} (\bibinfo {year} {2004})}\BibitemShut {NoStop}%
\bibitem [{\citenamefont {Buttinoni}\ \emph {et~al.}(2012)\citenamefont
  {Buttinoni}, \citenamefont {Volpe}, \citenamefont {K{\"u}mmel}, \citenamefont
  {Volpe},\ and\ \citenamefont {Bechinger}}]{buttinoni2012active}%
  \BibitemOpen
  \bibfield  {author} {\bibinfo {author} {\bibfnamefont {I.}~\bibnamefont
  {Buttinoni}}, \bibinfo {author} {\bibfnamefont {G.}~\bibnamefont {Volpe}},
  \bibinfo {author} {\bibfnamefont {F.}~\bibnamefont {K{\"u}mmel}}, \bibinfo
  {author} {\bibfnamefont {G.}~\bibnamefont {Volpe}}, \ and\ \bibinfo {author}
  {\bibfnamefont {C.}~\bibnamefont {Bechinger}},\ }\href@noop {} {\bibfield
  {journal} {\bibinfo  {journal} {J. Phys.: Condens. Matter}\ }\textbf
  {\bibinfo {volume} {24}},\ \bibinfo {pages} {284129} (\bibinfo {year}
  {2012})}\BibitemShut {NoStop}%
\bibitem [{\citenamefont {Walther}\ and\ \citenamefont
  {M\"uller}(2013)}]{walther2013janus}%
  \BibitemOpen
  \bibfield  {author} {\bibinfo {author} {\bibfnamefont {A.}~\bibnamefont
  {Walther}}\ and\ \bibinfo {author} {\bibfnamefont {A.~H.}\ \bibnamefont
  {M\"uller}},\ }\href@noop {} {\bibfield  {journal} {\bibinfo  {journal}
  {Chem. Rev.}\ }\textbf {\bibinfo {volume} {113}},\ \bibinfo {pages} {5194}
  (\bibinfo {year} {2013})}\BibitemShut {NoStop}%
\bibitem [{\citenamefont {Samin}\ and\ \citenamefont {van
  Roij}(2015)}]{samin2015self}%
  \BibitemOpen
  \bibfield  {author} {\bibinfo {author} {\bibfnamefont {S.}~\bibnamefont
  {Samin}}\ and\ \bibinfo {author} {\bibfnamefont {R.}~\bibnamefont {van
  Roij}},\ }\href@noop {} {\bibfield  {journal} {\bibinfo  {journal} {Phys.
  Rev. Lett.}\ }\textbf {\bibinfo {volume} {115}},\ \bibinfo {pages} {188305}
  (\bibinfo {year} {2015})}\BibitemShut {NoStop}%
\bibitem [{\citenamefont {Ramaswamy}(2010)}]{ramaswamy2010mechanics}%
  \BibitemOpen
  \bibfield  {author} {\bibinfo {author} {\bibfnamefont {S.}~\bibnamefont
  {Ramaswamy}},\ }\href@noop {} {\bibfield  {journal} {\bibinfo  {journal}
  {Annu. Rev. Condens. Matter Phys.}\ }\textbf {\bibinfo {volume} {1}},\
  \bibinfo {pages} {323} (\bibinfo {year} {2010})}\BibitemShut {NoStop}%
\bibitem [{\citenamefont {Marchetti}\ \emph {et~al.}(2013)\citenamefont
  {Marchetti}, \citenamefont {Joanny}, \citenamefont {Ramaswamy}, \citenamefont
  {Liverpool}, \citenamefont {Prost}, \citenamefont {Rao},\ and\ \citenamefont
  {Simha}}]{marchetti2013hydrodynamics}%
  \BibitemOpen
  \bibfield  {author} {\bibinfo {author} {\bibfnamefont {M.~C.}\ \bibnamefont
  {Marchetti}}, \bibinfo {author} {\bibfnamefont {J.-F.}\ \bibnamefont
  {Joanny}}, \bibinfo {author} {\bibfnamefont {S.}~\bibnamefont {Ramaswamy}},
  \bibinfo {author} {\bibfnamefont {T.~B.}\ \bibnamefont {Liverpool}}, \bibinfo
  {author} {\bibfnamefont {J.}~\bibnamefont {Prost}}, \bibinfo {author}
  {\bibfnamefont {M.}~\bibnamefont {Rao}}, \ and\ \bibinfo {author}
  {\bibfnamefont {R.~A.}\ \bibnamefont {Simha}},\ }\href@noop {} {\bibfield
  {journal} {\bibinfo  {journal} {Rev. Mod. Phys.}\ }\textbf {\bibinfo {volume}
  {85}},\ \bibinfo {pages} {1143} (\bibinfo {year} {2013})}\BibitemShut
  {NoStop}%
\bibitem [{\citenamefont {Menzel}(2015)}]{menzel2015tuned}%
  \BibitemOpen
  \bibfield  {author} {\bibinfo {author} {\bibfnamefont {A.~M.}\ \bibnamefont
  {Menzel}},\ }\href@noop {} {\bibfield  {journal} {\bibinfo  {journal} {Phys.
  Rep.}\ }\textbf {\bibinfo {volume} {554}},\ \bibinfo {pages} {1} (\bibinfo
  {year} {2015})}\BibitemShut {NoStop}%
\bibitem [{\citenamefont {Cates}\ and\ \citenamefont
  {Tailleur}(2013)}]{cates2013active}%
  \BibitemOpen
  \bibfield  {author} {\bibinfo {author} {\bibfnamefont {M.~E.}\ \bibnamefont
  {Cates}}\ and\ \bibinfo {author} {\bibfnamefont {J.}~\bibnamefont
  {Tailleur}},\ }\href@noop {} {\bibfield  {journal} {\bibinfo  {journal} {EPL
  (Europhys. Lett.)}\ }\textbf {\bibinfo {volume} {101}},\ \bibinfo {pages}
  {20010} (\bibinfo {year} {2013})}\BibitemShut {NoStop}%
\bibitem [{\citenamefont {Buttinoni}\ \emph {et~al.}(2013)\citenamefont
  {Buttinoni}, \citenamefont {Bialk{\'e}}, \citenamefont {K{\"u}mmel},
  \citenamefont {L{\"o}wen}, \citenamefont {Bechinger},\ and\ \citenamefont
  {Speck}}]{buttinoni2013dynamical}%
  \BibitemOpen
  \bibfield  {author} {\bibinfo {author} {\bibfnamefont {I.}~\bibnamefont
  {Buttinoni}}, \bibinfo {author} {\bibfnamefont {J.}~\bibnamefont
  {Bialk{\'e}}}, \bibinfo {author} {\bibfnamefont {F.}~\bibnamefont
  {K{\"u}mmel}}, \bibinfo {author} {\bibfnamefont {H.}~\bibnamefont
  {L{\"o}wen}}, \bibinfo {author} {\bibfnamefont {C.}~\bibnamefont
  {Bechinger}}, \ and\ \bibinfo {author} {\bibfnamefont {T.}~\bibnamefont
  {Speck}},\ }\href@noop {} {\bibfield  {journal} {\bibinfo  {journal} {Phys.
  Rev. Lett.}\ }\textbf {\bibinfo {volume} {110}},\ \bibinfo {pages} {238301}
  (\bibinfo {year} {2013})}\BibitemShut {NoStop}%
\bibitem [{\citenamefont {Bialk{\'e}}, \citenamefont {L{\"o}wen},\ and\
  \citenamefont {Speck}(2013)}]{bialke2013microscopic}%
  \BibitemOpen
  \bibfield  {author} {\bibinfo {author} {\bibfnamefont {J.}~\bibnamefont
  {Bialk{\'e}}}, \bibinfo {author} {\bibfnamefont {H.}~\bibnamefont
  {L{\"o}wen}}, \ and\ \bibinfo {author} {\bibfnamefont {T.}~\bibnamefont
  {Speck}},\ }\href@noop {} {\bibfield  {journal} {\bibinfo  {journal} {EPL
  (Europhys. Lett.)}\ }\textbf {\bibinfo {volume} {103}},\ \bibinfo {pages}
  {30008} (\bibinfo {year} {2013})}\BibitemShut {NoStop}%
\bibitem [{\citenamefont {Bialk{\'e}}, \citenamefont {Speck},\ and\
  \citenamefont {L{\"o}wen}(2015)}]{bialke2015active}%
  \BibitemOpen
  \bibfield  {author} {\bibinfo {author} {\bibfnamefont {J.}~\bibnamefont
  {Bialk{\'e}}}, \bibinfo {author} {\bibfnamefont {T.}~\bibnamefont {Speck}}, \
  and\ \bibinfo {author} {\bibfnamefont {H.}~\bibnamefont {L{\"o}wen}},\
  }\href@noop {} {\bibfield  {journal} {\bibinfo  {journal} {J. Non-Cryst.
  Solids}\ }\textbf {\bibinfo {volume} {407}},\ \bibinfo {pages} {367}
  (\bibinfo {year} {2015})}\BibitemShut {NoStop}%
\bibitem [{\citenamefont {Cates}\ and\ \citenamefont
  {Tailleur}(2015)}]{cates2015motility}%
  \BibitemOpen
  \bibfield  {author} {\bibinfo {author} {\bibfnamefont {M.~E.}\ \bibnamefont
  {Cates}}\ and\ \bibinfo {author} {\bibfnamefont {J.}~\bibnamefont
  {Tailleur}},\ }\href@noop {} {\bibfield  {journal} {\bibinfo  {journal}
  {Annu. Rev. Condens. Matter Phys.}\ }\textbf {\bibinfo {volume} {6}},\
  \bibinfo {pages} {219} (\bibinfo {year} {2015})}\BibitemShut {NoStop}%
\bibitem [{\citenamefont {Wittkowski}, \citenamefont {Stenhammar},\ and\
  \citenamefont {Cates}(2017)}]{wittkowski2017nonequilibrium}%
  \BibitemOpen
  \bibfield  {author} {\bibinfo {author} {\bibfnamefont {R.}~\bibnamefont
  {Wittkowski}}, \bibinfo {author} {\bibfnamefont {J.}~\bibnamefont
  {Stenhammar}}, \ and\ \bibinfo {author} {\bibfnamefont {M.~E.}\ \bibnamefont
  {Cates}},\ }\href@noop {} {\bibfield  {journal} {\bibinfo  {journal} {New J.
  Phys.}\ }\textbf {\bibinfo {volume} {19}},\ \bibinfo {pages} {105003}
  (\bibinfo {year} {2017})}\BibitemShut {NoStop}%
\bibitem [{\citenamefont {Solon}\ \emph {et~al.}(2018)\citenamefont {Solon},
  \citenamefont {Stenhammar}, \citenamefont {Cates}, \citenamefont {Kafri},\
  and\ \citenamefont {Tailleur}}]{solon2018generalized_NJP}%
  \BibitemOpen
  \bibfield  {author} {\bibinfo {author} {\bibfnamefont {A.}~\bibnamefont
  {Solon}}, \bibinfo {author} {\bibfnamefont {J.}~\bibnamefont {Stenhammar}},
  \bibinfo {author} {\bibfnamefont {M.~E.}\ \bibnamefont {Cates}}, \bibinfo
  {author} {\bibfnamefont {Y.}~\bibnamefont {Kafri}}, \ and\ \bibinfo {author}
  {\bibfnamefont {J.}~\bibnamefont {Tailleur}},\ }\href@noop {} {\bibfield
  {journal} {\bibinfo  {journal} {New J. Phys.}\ }\textbf {\bibinfo {volume}
  {20}},\ \bibinfo {pages} {075001} (\bibinfo {year} {2018})}\BibitemShut
  {NoStop}%
\bibitem [{\citenamefont {Digregorio}\ \emph {et~al.}(2018)\citenamefont
  {Digregorio}, \citenamefont {Levis}, \citenamefont {Suma}, \citenamefont
  {Cugliandolo}, \citenamefont {Gonnella},\ and\ \citenamefont
  {Pagonabarraga}}]{digregorio2018full}%
  \BibitemOpen
  \bibfield  {author} {\bibinfo {author} {\bibfnamefont {P.}~\bibnamefont
  {Digregorio}}, \bibinfo {author} {\bibfnamefont {D.}~\bibnamefont {Levis}},
  \bibinfo {author} {\bibfnamefont {A.}~\bibnamefont {Suma}}, \bibinfo {author}
  {\bibfnamefont {L.~F.}\ \bibnamefont {Cugliandolo}}, \bibinfo {author}
  {\bibfnamefont {G.}~\bibnamefont {Gonnella}}, \ and\ \bibinfo {author}
  {\bibfnamefont {I.}~\bibnamefont {Pagonabarraga}},\ }\href@noop {} {\bibfield
   {journal} {\bibinfo  {journal} {preprint arXiv:1805.12484}\ } (\bibinfo
  {year} {2018})}\BibitemShut {NoStop}%
\bibitem [{\citenamefont {Wensink}\ and\ \citenamefont
  {L{\"o}wen}(2012)}]{wensink2012emergent}%
  \BibitemOpen
  \bibfield  {author} {\bibinfo {author} {\bibfnamefont {H.}~\bibnamefont
  {Wensink}}\ and\ \bibinfo {author} {\bibfnamefont {H.}~\bibnamefont
  {L{\"o}wen}},\ }\href@noop {} {\bibfield  {journal} {\bibinfo  {journal} {J.
  Phys.: Condens. Matter}\ }\textbf {\bibinfo {volume} {24}},\ \bibinfo {pages}
  {464130} (\bibinfo {year} {2012})}\BibitemShut {NoStop}%
\bibitem [{\citenamefont {Menzel}(2013)}]{menzel2013unidirectional}%
  \BibitemOpen
  \bibfield  {author} {\bibinfo {author} {\bibfnamefont {A.~M.}\ \bibnamefont
  {Menzel}},\ }\href@noop {} {\bibfield  {journal} {\bibinfo  {journal} {J.
  Phys.: Condens. Matter}\ }\textbf {\bibinfo {volume} {25}},\ \bibinfo {pages}
  {505103} (\bibinfo {year} {2013})}\BibitemShut {NoStop}%
\bibitem [{\citenamefont {Kogler}\ and\ \citenamefont
  {Klapp}(2015)}]{kogler2015lane}%
  \BibitemOpen
  \bibfield  {author} {\bibinfo {author} {\bibfnamefont {F.}~\bibnamefont
  {Kogler}}\ and\ \bibinfo {author} {\bibfnamefont {S.~H.~L.}\ \bibnamefont
  {Klapp}},\ }\href@noop {} {\bibfield  {journal} {\bibinfo  {journal} {EPL
  (Europhys. Lett.)}\ }\textbf {\bibinfo {volume} {110}},\ \bibinfo {pages}
  {10004} (\bibinfo {year} {2015})}\BibitemShut {NoStop}%
\bibitem [{\citenamefont {Romanczuk}\ \emph {et~al.}(2016)\citenamefont
  {Romanczuk}, \citenamefont {Chat{\'e}}, \citenamefont {Chen}, \citenamefont
  {Ngo},\ and\ \citenamefont {Toner}}]{romanczuk2016emergent}%
  \BibitemOpen
  \bibfield  {author} {\bibinfo {author} {\bibfnamefont {P.}~\bibnamefont
  {Romanczuk}}, \bibinfo {author} {\bibfnamefont {H.}~\bibnamefont
  {Chat{\'e}}}, \bibinfo {author} {\bibfnamefont {L.}~\bibnamefont {Chen}},
  \bibinfo {author} {\bibfnamefont {S.}~\bibnamefont {Ngo}}, \ and\ \bibinfo
  {author} {\bibfnamefont {J.}~\bibnamefont {Toner}},\ }\href@noop {}
  {\bibfield  {journal} {\bibinfo  {journal} {New J. Phys.}\ }\textbf {\bibinfo
  {volume} {18}},\ \bibinfo {pages} {063015} (\bibinfo {year}
  {2016})}\BibitemShut {NoStop}%
\bibitem [{\citenamefont {Menzel}(2016)}]{menzel2016way}%
  \BibitemOpen
  \bibfield  {author} {\bibinfo {author} {\bibfnamefont {A.~M.}\ \bibnamefont
  {Menzel}},\ }\href@noop {} {\bibfield  {journal} {\bibinfo  {journal} {New J.
  Phys.}\ }\textbf {\bibinfo {volume} {18}},\ \bibinfo {pages} {071001}
  (\bibinfo {year} {2016})}\BibitemShut {NoStop}%
\bibitem [{\citenamefont {Dusenbery}(2009)}]{dusenbery2009living}%
  \BibitemOpen
  \bibfield  {author} {\bibinfo {author} {\bibfnamefont {D.~B.}\ \bibnamefont
  {Dusenbery}},\ }\href@noop {} {\emph {\bibinfo {title} {Living at Micro
  Scale:\ The Unexpected Physics of Being Small}}}\ (\bibinfo  {publisher}
  {Harvard University Press, Cambridge},\ \bibinfo {year} {2009})\BibitemShut
  {NoStop}%
\bibitem [{\citenamefont {Kessler}(1985)}]{kessler1985hydrodynamic}%
  \BibitemOpen
  \bibfield  {author} {\bibinfo {author} {\bibfnamefont {J.~O.}\ \bibnamefont
  {Kessler}},\ }\href@noop {} {\bibfield  {journal} {\bibinfo  {journal}
  {Nature}\ }\textbf {\bibinfo {volume} {313}},\ \bibinfo {pages} {218}
  (\bibinfo {year} {1985})}\BibitemShut {NoStop}%
\bibitem [{\citenamefont {Durham}, \citenamefont {Kessler},\ and\ \citenamefont
  {Stocker}(2009)}]{durham2009disruption}%
  \BibitemOpen
  \bibfield  {author} {\bibinfo {author} {\bibfnamefont {W.~M.}\ \bibnamefont
  {Durham}}, \bibinfo {author} {\bibfnamefont {J.~O.}\ \bibnamefont {Kessler}},
  \ and\ \bibinfo {author} {\bibfnamefont {R.}~\bibnamefont {Stocker}},\
  }\href@noop {} {\bibfield  {journal} {\bibinfo  {journal} {Science}\ }\textbf
  {\bibinfo {volume} {323}},\ \bibinfo {pages} {1067} (\bibinfo {year}
  {2009})}\BibitemShut {NoStop}%
\bibitem [{\citenamefont {ten Hagen}\ \emph {et~al.}(2014)\citenamefont {ten
  Hagen}, \citenamefont {K{\"u}mmel}, \citenamefont {Wittkowski}, \citenamefont
  {Takagi}, \citenamefont {L{\"o}wen},\ and\ \citenamefont
  {Bechinger}}]{hagen2014gravitaxis}%
  \BibitemOpen
  \bibfield  {author} {\bibinfo {author} {\bibfnamefont {B.}~\bibnamefont {ten
  Hagen}}, \bibinfo {author} {\bibfnamefont {F.}~\bibnamefont {K{\"u}mmel}},
  \bibinfo {author} {\bibfnamefont {R.}~\bibnamefont {Wittkowski}}, \bibinfo
  {author} {\bibfnamefont {D.}~\bibnamefont {Takagi}}, \bibinfo {author}
  {\bibfnamefont {H.}~\bibnamefont {L{\"o}wen}}, \ and\ \bibinfo {author}
  {\bibfnamefont {C.}~\bibnamefont {Bechinger}},\ }\href {\doibase
  10.1038/ncomms5829} {\bibfield  {journal} {\bibinfo  {journal} {Nat.
  Commun.}\ }\textbf {\bibinfo {volume} {5}},\ \bibinfo {pages} {4829}
  (\bibinfo {year} {2014})}\BibitemShut {NoStop}%
\bibitem [{\citenamefont {Lozano}\ \emph {et~al.}(2016)\citenamefont {Lozano},
  \citenamefont {ten Hagen}, \citenamefont {L{\"o}wen},\ and\ \citenamefont
  {Bechinger}}]{lozano2016phototaxis}%
  \BibitemOpen
  \bibfield  {author} {\bibinfo {author} {\bibfnamefont {C.}~\bibnamefont
  {Lozano}}, \bibinfo {author} {\bibfnamefont {B.}~\bibnamefont {ten Hagen}},
  \bibinfo {author} {\bibfnamefont {H.}~\bibnamefont {L{\"o}wen}}, \ and\
  \bibinfo {author} {\bibfnamefont {C.}~\bibnamefont {Bechinger}},\ }\href@noop
  {} {\bibfield  {journal} {\bibinfo  {journal} {Nat. Commun.}\ }\textbf
  {\bibinfo {volume} {7}},\ \bibinfo {pages} {12828} (\bibinfo {year}
  {2016})}\BibitemShut {NoStop}%
\bibitem [{\citenamefont {Campbell}\ \emph {et~al.}(2017)\citenamefont
  {Campbell}, \citenamefont {Wittkowski}, \citenamefont {ten Hagen},
  \citenamefont {L{\"o}wen},\ and\ \citenamefont
  {Ebbens}}]{campbell2017helical}%
  \BibitemOpen
  \bibfield  {author} {\bibinfo {author} {\bibfnamefont {A.~I.}\ \bibnamefont
  {Campbell}}, \bibinfo {author} {\bibfnamefont {R.}~\bibnamefont
  {Wittkowski}}, \bibinfo {author} {\bibfnamefont {B.}~\bibnamefont {ten
  Hagen}}, \bibinfo {author} {\bibfnamefont {H.}~\bibnamefont {L{\"o}wen}}, \
  and\ \bibinfo {author} {\bibfnamefont {S.~J.}\ \bibnamefont {Ebbens}},\
  }\href@noop {} {\bibfield  {journal} {\bibinfo  {journal} {J. Chem. Phys.}\
  }\textbf {\bibinfo {volume} {147}},\ \bibinfo {pages} {084905} (\bibinfo
  {year} {2017})}\BibitemShut {NoStop}%
\bibitem [{\citenamefont {Liebchen}\ \emph {et~al.}(2018)\citenamefont
  {Liebchen}, \citenamefont {Monderkamp}, \citenamefont {ten Hagen},\ and\
  \citenamefont {L\"owen}}]{liebchen2018viscotaxis}%
  \BibitemOpen
  \bibfield  {author} {\bibinfo {author} {\bibfnamefont {B.}~\bibnamefont
  {Liebchen}}, \bibinfo {author} {\bibfnamefont {P.}~\bibnamefont
  {Monderkamp}}, \bibinfo {author} {\bibfnamefont {B.}~\bibnamefont {ten
  Hagen}}, \ and\ \bibinfo {author} {\bibfnamefont {H.}~\bibnamefont
  {L\"owen}},\ }\href@noop {} {\bibfield  {journal} {\bibinfo  {journal} {Phys.
  Rev. Lett.}\ }\textbf {\bibinfo {volume} {120}},\ \bibinfo {pages} {208002}
  (\bibinfo {year} {2018})}\BibitemShut {NoStop}%
\bibitem [{\citenamefont {Wensink}\ \emph {et~al.}(2012)\citenamefont
  {Wensink}, \citenamefont {Dunkel}, \citenamefont {Heidenreich}, \citenamefont
  {Drescher}, \citenamefont {Goldstein}, \citenamefont {L{\"o}wen},\ and\
  \citenamefont {Yeomans}}]{wensink2012meso}%
  \BibitemOpen
  \bibfield  {author} {\bibinfo {author} {\bibfnamefont {H.~H.}\ \bibnamefont
  {Wensink}}, \bibinfo {author} {\bibfnamefont {J.}~\bibnamefont {Dunkel}},
  \bibinfo {author} {\bibfnamefont {S.}~\bibnamefont {Heidenreich}}, \bibinfo
  {author} {\bibfnamefont {K.}~\bibnamefont {Drescher}}, \bibinfo {author}
  {\bibfnamefont {R.~E.}\ \bibnamefont {Goldstein}}, \bibinfo {author}
  {\bibfnamefont {H.}~\bibnamefont {L{\"o}wen}}, \ and\ \bibinfo {author}
  {\bibfnamefont {J.~M.}\ \bibnamefont {Yeomans}},\ }\href@noop {} {\bibfield
  {journal} {\bibinfo  {journal} {Proc. Natl. Acad. Sci. U.S.A.}\ }\textbf
  {\bibinfo {volume} {109}},\ \bibinfo {pages} {14308} (\bibinfo {year}
  {2012})}\BibitemShut {NoStop}%
\bibitem [{\citenamefont {Sokolov}\ and\ \citenamefont
  {Aranson}(2012)}]{sokolov2012physical}%
  \BibitemOpen
  \bibfield  {author} {\bibinfo {author} {\bibfnamefont {A.}~\bibnamefont
  {Sokolov}}\ and\ \bibinfo {author} {\bibfnamefont {I.~S.}\ \bibnamefont
  {Aranson}},\ }\href@noop {} {\bibfield  {journal} {\bibinfo  {journal} {Phys.
  Rev. Lett.}\ }\textbf {\bibinfo {volume} {109}},\ \bibinfo {pages} {248109}
  (\bibinfo {year} {2012})}\BibitemShut {NoStop}%
\bibitem [{\citenamefont {Dunkel}\ \emph {et~al.}(2013)\citenamefont {Dunkel},
  \citenamefont {Heidenreich}, \citenamefont {Drescher}, \citenamefont
  {Wensink}, \citenamefont {B{\"a}r},\ and\ \citenamefont
  {Goldstein}}]{dunkel2013fluid}%
  \BibitemOpen
  \bibfield  {author} {\bibinfo {author} {\bibfnamefont {J.}~\bibnamefont
  {Dunkel}}, \bibinfo {author} {\bibfnamefont {S.}~\bibnamefont {Heidenreich}},
  \bibinfo {author} {\bibfnamefont {K.}~\bibnamefont {Drescher}}, \bibinfo
  {author} {\bibfnamefont {H.~H.}\ \bibnamefont {Wensink}}, \bibinfo {author}
  {\bibfnamefont {M.}~\bibnamefont {B{\"a}r}}, \ and\ \bibinfo {author}
  {\bibfnamefont {R.~E.}\ \bibnamefont {Goldstein}},\ }\href@noop {} {\bibfield
   {journal} {\bibinfo  {journal} {Phys. Rev. Lett.}\ }\textbf {\bibinfo
  {volume} {110}},\ \bibinfo {pages} {228102} (\bibinfo {year}
  {2013})}\BibitemShut {NoStop}%
\bibitem [{\citenamefont {Kaiser}\ \emph {et~al.}(2014)\citenamefont {Kaiser},
  \citenamefont {Peshkov}, \citenamefont {Sokolov}, \citenamefont {ten Hagen},
  \citenamefont {L{\"o}wen},\ and\ \citenamefont
  {Aranson}}]{kaiser2014transport}%
  \BibitemOpen
  \bibfield  {author} {\bibinfo {author} {\bibfnamefont {A.}~\bibnamefont
  {Kaiser}}, \bibinfo {author} {\bibfnamefont {A.}~\bibnamefont {Peshkov}},
  \bibinfo {author} {\bibfnamefont {A.}~\bibnamefont {Sokolov}}, \bibinfo
  {author} {\bibfnamefont {B.}~\bibnamefont {ten Hagen}}, \bibinfo {author}
  {\bibfnamefont {H.}~\bibnamefont {L{\"o}wen}}, \ and\ \bibinfo {author}
  {\bibfnamefont {I.~S.}\ \bibnamefont {Aranson}},\ }\href@noop {} {\bibfield
  {journal} {\bibinfo  {journal} {Phys. Rev. Lett.}\ }\textbf {\bibinfo
  {volume} {112}},\ \bibinfo {pages} {158101} (\bibinfo {year}
  {2014})}\BibitemShut {NoStop}%
\bibitem [{\citenamefont {S{\l}omka}\ and\ \citenamefont
  {Dunkel}(2015)}]{slomka2015generalized}%
  \BibitemOpen
  \bibfield  {author} {\bibinfo {author} {\bibfnamefont {J.}~\bibnamefont
  {S{\l}omka}}\ and\ \bibinfo {author} {\bibfnamefont {J.}~\bibnamefont
  {Dunkel}},\ }\href@noop {} {\bibfield  {journal} {\bibinfo  {journal} {Eur.
  Phys. J.: Spec. Top.}\ }\textbf {\bibinfo {volume} {224}},\ \bibinfo {pages}
  {1349} (\bibinfo {year} {2015})}\BibitemShut {NoStop}%
\bibitem [{\citenamefont {Cates}(2012)}]{cates2012diffusive}%
  \BibitemOpen
  \bibfield  {author} {\bibinfo {author} {\bibfnamefont {M.~E.}\ \bibnamefont
  {Cates}},\ }\href {\doibase 10.1088/0034-4885/75/4/042601} {\bibfield
  {journal} {\bibinfo  {journal} {Rep. Prog. Phys.}\ }\textbf {\bibinfo
  {volume} {75}},\ \bibinfo {pages} {042601} (\bibinfo {year}
  {2012})}\BibitemShut {NoStop}%
\bibitem [{\citenamefont {ten Hagen}\ \emph {et~al.}(2015)\citenamefont {ten
  Hagen}, \citenamefont {Wittkowski}, \citenamefont {Takagi}, \citenamefont
  {K{\"u}mmel}, \citenamefont {Bechinger},\ and\ \citenamefont
  {L{\"o}wen}}]{ten2015can}%
  \BibitemOpen
  \bibfield  {author} {\bibinfo {author} {\bibfnamefont {B.}~\bibnamefont {ten
  Hagen}}, \bibinfo {author} {\bibfnamefont {R.}~\bibnamefont {Wittkowski}},
  \bibinfo {author} {\bibfnamefont {D.}~\bibnamefont {Takagi}}, \bibinfo
  {author} {\bibfnamefont {F.}~\bibnamefont {K{\"u}mmel}}, \bibinfo {author}
  {\bibfnamefont {C.}~\bibnamefont {Bechinger}}, \ and\ \bibinfo {author}
  {\bibfnamefont {H.}~\bibnamefont {L{\"o}wen}},\ }\href@noop {} {\bibfield
  {journal} {\bibinfo  {journal} {J. Phys.: Condens. Matter}\ }\textbf
  {\bibinfo {volume} {27}},\ \bibinfo {pages} {194110} (\bibinfo {year}
  {2015})}\BibitemShut {NoStop}%
\bibitem [{\citenamefont {Yan}\ and\ \citenamefont
  {Brady}(2015{\natexlab{a}})}]{yan2015force}%
  \BibitemOpen
  \bibfield  {author} {\bibinfo {author} {\bibfnamefont {W.}~\bibnamefont
  {Yan}}\ and\ \bibinfo {author} {\bibfnamefont {J.~F.}\ \bibnamefont
  {Brady}},\ }\href@noop {} {\bibfield  {journal} {\bibinfo  {journal} {J.
  Fluid Mech.}\ }\textbf {\bibinfo {volume} {785}},\ \bibinfo {pages} {R1}
  (\bibinfo {year} {2015}{\natexlab{a}})}\BibitemShut {NoStop}%
\bibitem [{\citenamefont {Yan}\ and\ \citenamefont
  {Brady}(2015{\natexlab{b}})}]{yan2015swim}%
  \BibitemOpen
  \bibfield  {author} {\bibinfo {author} {\bibfnamefont {W.}~\bibnamefont
  {Yan}}\ and\ \bibinfo {author} {\bibfnamefont {J.~F.}\ \bibnamefont
  {Brady}},\ }\href@noop {} {\bibfield  {journal} {\bibinfo  {journal} {Soft
  Matter}\ }\textbf {\bibinfo {volume} {11}},\ \bibinfo {pages} {6235}
  (\bibinfo {year} {2015}{\natexlab{b}})}\BibitemShut {NoStop}%
\bibitem [{\citenamefont {Takatori}, \citenamefont {Yan},\ and\ \citenamefont
  {Brady}(2014)}]{takatori2014swim}%
  \BibitemOpen
  \bibfield  {author} {\bibinfo {author} {\bibfnamefont {S.~C.}\ \bibnamefont
  {Takatori}}, \bibinfo {author} {\bibfnamefont {W.}~\bibnamefont {Yan}}, \
  and\ \bibinfo {author} {\bibfnamefont {J.~F.}\ \bibnamefont {Brady}},\
  }\href@noop {} {\bibfield  {journal} {\bibinfo  {journal} {Phys. Rev. Lett.}\
  }\textbf {\bibinfo {volume} {113}},\ \bibinfo {pages} {028103} (\bibinfo
  {year} {2014})}\BibitemShut {NoStop}%
\bibitem [{\citenamefont {Solon}\ \emph {et~al.}(2015)\citenamefont {Solon},
  \citenamefont {Fily}, \citenamefont {Baskaran}, \citenamefont {Cates},
  \citenamefont {Kafri}, \citenamefont {Kardar},\ and\ \citenamefont
  {Tailleur}}]{solon2015pressure}%
  \BibitemOpen
  \bibfield  {author} {\bibinfo {author} {\bibfnamefont {A.~P.}\ \bibnamefont
  {Solon}}, \bibinfo {author} {\bibfnamefont {Y.}~\bibnamefont {Fily}},
  \bibinfo {author} {\bibfnamefont {A.}~\bibnamefont {Baskaran}}, \bibinfo
  {author} {\bibfnamefont {M.~E.}\ \bibnamefont {Cates}}, \bibinfo {author}
  {\bibfnamefont {Y.}~\bibnamefont {Kafri}}, \bibinfo {author} {\bibfnamefont
  {M.}~\bibnamefont {Kardar}}, \ and\ \bibinfo {author} {\bibfnamefont
  {J.}~\bibnamefont {Tailleur}},\ }\href@noop {} {\bibfield  {journal}
  {\bibinfo  {journal} {Nat. Phys.}\ }\textbf {\bibinfo {volume} {11}},\
  \bibinfo {pages} {673} (\bibinfo {year} {2015})}\BibitemShut {NoStop}%
\bibitem [{\citenamefont {Smallenburg}\ and\ \citenamefont
  {L{\"o}wen}(2015)}]{smallenburg2015swim}%
  \BibitemOpen
  \bibfield  {author} {\bibinfo {author} {\bibfnamefont {F.}~\bibnamefont
  {Smallenburg}}\ and\ \bibinfo {author} {\bibfnamefont {H.}~\bibnamefont
  {L{\"o}wen}},\ }\href@noop {} {\bibfield  {journal} {\bibinfo  {journal}
  {Phys. Rev. E}\ }\textbf {\bibinfo {volume} {92}},\ \bibinfo {pages} {032304}
  (\bibinfo {year} {2015})}\BibitemShut {NoStop}%
\bibitem [{\citenamefont {Liluashvili}, \citenamefont {{\'O}nody},\ and\
  \citenamefont {Voigtmann}(2017)}]{liluashvili2017mode}%
  \BibitemOpen
  \bibfield  {author} {\bibinfo {author} {\bibfnamefont {A.}~\bibnamefont
  {Liluashvili}}, \bibinfo {author} {\bibfnamefont {J.}~\bibnamefont
  {{\'O}nody}}, \ and\ \bibinfo {author} {\bibfnamefont {T.}~\bibnamefont
  {Voigtmann}},\ }\href@noop {} {\bibfield  {journal} {\bibinfo  {journal}
  {Phys. Rev. E}\ }\textbf {\bibinfo {volume} {96}},\ \bibinfo {pages} {062608}
  (\bibinfo {year} {2017})}\BibitemShut {NoStop}%
\bibitem [{\citenamefont {Engstler}\ \emph {et~al.}(2007)\citenamefont
  {Engstler}, \citenamefont {Pfohl}, \citenamefont {Herminghaus}, \citenamefont
  {Boshart}, \citenamefont {Wiegertjes}, \citenamefont {Heddergott},\ and\
  \citenamefont {Overath}}]{engstler2007hydrodynamic}%
  \BibitemOpen
  \bibfield  {author} {\bibinfo {author} {\bibfnamefont {M.}~\bibnamefont
  {Engstler}}, \bibinfo {author} {\bibfnamefont {T.}~\bibnamefont {Pfohl}},
  \bibinfo {author} {\bibfnamefont {S.}~\bibnamefont {Herminghaus}}, \bibinfo
  {author} {\bibfnamefont {M.}~\bibnamefont {Boshart}}, \bibinfo {author}
  {\bibfnamefont {G.}~\bibnamefont {Wiegertjes}}, \bibinfo {author}
  {\bibfnamefont {N.}~\bibnamefont {Heddergott}}, \ and\ \bibinfo {author}
  {\bibfnamefont {P.}~\bibnamefont {Overath}},\ }\href@noop {} {\bibfield
  {journal} {\bibinfo  {journal} {Cell}\ }\textbf {\bibinfo {volume} {131}},\
  \bibinfo {pages} {505} (\bibinfo {year} {2007})}\BibitemShut {NoStop}%
\bibitem [{\citenamefont {Nelson}, \citenamefont {Kaliakatsos},\ and\
  \citenamefont {Abbott}(2010)}]{nelson2010microrobots}%
  \BibitemOpen
  \bibfield  {author} {\bibinfo {author} {\bibfnamefont {B.~J.}\ \bibnamefont
  {Nelson}}, \bibinfo {author} {\bibfnamefont {I.~K.}\ \bibnamefont
  {Kaliakatsos}}, \ and\ \bibinfo {author} {\bibfnamefont {J.~J.}\ \bibnamefont
  {Abbott}},\ }\href@noop {} {\bibfield  {journal} {\bibinfo  {journal} {Annu.
  Rev. Biomed. Eng.}\ }\textbf {\bibinfo {volume} {12}},\ \bibinfo {pages} {55}
  (\bibinfo {year} {2010})}\BibitemShut {NoStop}%
\bibitem [{\citenamefont {Wang}\ and\ \citenamefont
  {Gao}(2012)}]{wang2012nano}%
  \BibitemOpen
  \bibfield  {author} {\bibinfo {author} {\bibfnamefont {J.}~\bibnamefont
  {Wang}}\ and\ \bibinfo {author} {\bibfnamefont {W.}~\bibnamefont {Gao}},\
  }\href@noop {} {\bibfield  {journal} {\bibinfo  {journal} {ACS Nano}\
  }\textbf {\bibinfo {volume} {6}},\ \bibinfo {pages} {5745} (\bibinfo {year}
  {2012})}\BibitemShut {NoStop}%
\bibitem [{\citenamefont {Patra}\ \emph {et~al.}(2013)\citenamefont {Patra},
  \citenamefont {Sengupta}, \citenamefont {Duan}, \citenamefont {Zhang},
  \citenamefont {Pavlick},\ and\ \citenamefont {Sen}}]{patra2013intelligent}%
  \BibitemOpen
  \bibfield  {author} {\bibinfo {author} {\bibfnamefont {D.}~\bibnamefont
  {Patra}}, \bibinfo {author} {\bibfnamefont {S.}~\bibnamefont {Sengupta}},
  \bibinfo {author} {\bibfnamefont {W.}~\bibnamefont {Duan}}, \bibinfo {author}
  {\bibfnamefont {H.}~\bibnamefont {Zhang}}, \bibinfo {author} {\bibfnamefont
  {R.}~\bibnamefont {Pavlick}}, \ and\ \bibinfo {author} {\bibfnamefont
  {A.}~\bibnamefont {Sen}},\ }\href@noop {} {\bibfield  {journal} {\bibinfo
  {journal} {Nanoscale}\ }\textbf {\bibinfo {volume} {5}},\ \bibinfo {pages}
  {1273} (\bibinfo {year} {2013})}\BibitemShut {NoStop}%
\bibitem [{\citenamefont {Xi}\ \emph {et~al.}(2013)\citenamefont {Xi},
  \citenamefont {Solovev}, \citenamefont {Ananth}, \citenamefont {Gracias},
  \citenamefont {Sanchez},\ and\ \citenamefont {Schmidt}}]{xi2013rolled}%
  \BibitemOpen
  \bibfield  {author} {\bibinfo {author} {\bibfnamefont {W.}~\bibnamefont
  {Xi}}, \bibinfo {author} {\bibfnamefont {A.~A.}\ \bibnamefont {Solovev}},
  \bibinfo {author} {\bibfnamefont {A.~N.}\ \bibnamefont {Ananth}}, \bibinfo
  {author} {\bibfnamefont {D.~H.}\ \bibnamefont {Gracias}}, \bibinfo {author}
  {\bibfnamefont {S.}~\bibnamefont {Sanchez}}, \ and\ \bibinfo {author}
  {\bibfnamefont {O.~G.}\ \bibnamefont {Schmidt}},\ }\href@noop {} {\bibfield
  {journal} {\bibinfo  {journal} {Nanoscale}\ }\textbf {\bibinfo {volume}
  {5}},\ \bibinfo {pages} {1294} (\bibinfo {year} {2013})}\BibitemShut
  {NoStop}%
\bibitem [{\citenamefont {Abdelmohsen}\ \emph {et~al.}(2014)\citenamefont
  {Abdelmohsen}, \citenamefont {Peng}, \citenamefont {Tu},\ and\ \citenamefont
  {Wilson}}]{abdelmohsen2014micro}%
  \BibitemOpen
  \bibfield  {author} {\bibinfo {author} {\bibfnamefont {L.~K.}\ \bibnamefont
  {Abdelmohsen}}, \bibinfo {author} {\bibfnamefont {F.}~\bibnamefont {Peng}},
  \bibinfo {author} {\bibfnamefont {Y.}~\bibnamefont {Tu}}, \ and\ \bibinfo
  {author} {\bibfnamefont {D.~A.}\ \bibnamefont {Wilson}},\ }\href@noop {}
  {\bibfield  {journal} {\bibinfo  {journal} {J. Mater. Chem. B}\ }\textbf
  {\bibinfo {volume} {2}},\ \bibinfo {pages} {2395} (\bibinfo {year}
  {2014})}\BibitemShut {NoStop}%
\bibitem [{\citenamefont {Dhont}(1996)}]{Dhont}%
  \BibitemOpen
  \bibfield  {author} {\bibinfo {author} {\bibfnamefont {J.~K.~G.}\
  \bibnamefont {Dhont}},\ }\href@noop {} {\emph {\bibinfo {title} {An
  Introduction to Dynamics of Colloids}}}\ (\bibinfo  {publisher} {Elsevier,
  Amsterdam},\ \bibinfo {year} {1996})\BibitemShut {NoStop}%
\bibitem [{\citenamefont {Happel}\ and\ \citenamefont
  {Brenner}(2012)}]{Happel_Brenner}%
  \BibitemOpen
  \bibfield  {author} {\bibinfo {author} {\bibfnamefont {J.}~\bibnamefont
  {Happel}}\ and\ \bibinfo {author} {\bibfnamefont {H.}~\bibnamefont
  {Brenner}},\ }\href@noop {} {\emph {\bibinfo {title} {Low Reynolds Number
  Hydrodynamics: With Special Applications to Particulate Media}}},\
  Vol.~\bibinfo {volume} {1}\ (\bibinfo  {publisher} {Springer Science \&
  Business Media, New York},\ \bibinfo {year} {2012})\BibitemShut {NoStop}%
\bibitem [{\citenamefont {Kim}\ and\ \citenamefont
  {Karrila}(2013)}]{Kim_Karrila}%
  \BibitemOpen
  \bibfield  {author} {\bibinfo {author} {\bibfnamefont {S.}~\bibnamefont
  {Kim}}\ and\ \bibinfo {author} {\bibfnamefont {S.~J.}\ \bibnamefont
  {Karrila}},\ }\href@noop {} {\emph {\bibinfo {title} {Microhydrodynamics:
  Principles and Selected Applications}}}\ (\bibinfo  {publisher} {Dover
  Publications, Mineola},\ \bibinfo {year} {2013})\BibitemShut {NoStop}%
\bibitem [{\citenamefont {Najafi}\ and\ \citenamefont
  {Golestanian}(2004)}]{najafi2004simple}%
  \BibitemOpen
  \bibfield  {author} {\bibinfo {author} {\bibfnamefont {A.}~\bibnamefont
  {Najafi}}\ and\ \bibinfo {author} {\bibfnamefont {R.}~\bibnamefont
  {Golestanian}},\ }\href@noop {} {\bibfield  {journal} {\bibinfo  {journal}
  {Phys. Rev. E}\ }\textbf {\bibinfo {volume} {69}},\ \bibinfo {pages} {062901}
  (\bibinfo {year} {2004})}\BibitemShut {NoStop}%
\bibitem [{\citenamefont {Zargar}, \citenamefont {Najafi},\ and\ \citenamefont
  {Miri}(2009)}]{zargar2009three}%
  \BibitemOpen
  \bibfield  {author} {\bibinfo {author} {\bibfnamefont {R.}~\bibnamefont
  {Zargar}}, \bibinfo {author} {\bibfnamefont {A.}~\bibnamefont {Najafi}}, \
  and\ \bibinfo {author} {\bibfnamefont {M.}~\bibnamefont {Miri}},\ }\href@noop
  {} {\bibfield  {journal} {\bibinfo  {journal} {Phys. Rev. E}\ }\textbf
  {\bibinfo {volume} {80}},\ \bibinfo {pages} {026308} (\bibinfo {year}
  {2009})}\BibitemShut {NoStop}%
\bibitem [{\citenamefont {Daddi-Moussa-Ider}\ \emph
  {et~al.}(2018{\natexlab{a}})\citenamefont {Daddi-Moussa-Ider}, \citenamefont
  {Lisicki}, \citenamefont {Hoell},\ and\ \citenamefont
  {L{\"o}wen}}]{daddi2018swimming}%
  \BibitemOpen
  \bibfield  {author} {\bibinfo {author} {\bibfnamefont {A.}~\bibnamefont
  {Daddi-Moussa-Ider}}, \bibinfo {author} {\bibfnamefont {M.}~\bibnamefont
  {Lisicki}}, \bibinfo {author} {\bibfnamefont {C.}~\bibnamefont {Hoell}}, \
  and\ \bibinfo {author} {\bibfnamefont {H.}~\bibnamefont {L{\"o}wen}},\
  }\href@noop {} {\bibfield  {journal} {\bibinfo  {journal} {J. Chem. Phys.}\
  }\textbf {\bibinfo {volume} {148}},\ \bibinfo {pages} {134904} (\bibinfo
  {year} {2018}{\natexlab{a}})}\BibitemShut {NoStop}%
\bibitem [{\citenamefont {Daddi-Moussa-Ider}\ \emph
  {et~al.}(2018{\natexlab{b}})\citenamefont {Daddi-Moussa-Ider}, \citenamefont
  {Lisicki}, \citenamefont {Mathijssen}, \citenamefont {Hoell}, \citenamefont
  {Goh}, \citenamefont {Blawzdziewicz}, \citenamefont {Menzel},\ and\
  \citenamefont {L{\"o}wen}}]{daddi2018state}%
  \BibitemOpen
  \bibfield  {author} {\bibinfo {author} {\bibfnamefont {A.}~\bibnamefont
  {Daddi-Moussa-Ider}}, \bibinfo {author} {\bibfnamefont {M.}~\bibnamefont
  {Lisicki}}, \bibinfo {author} {\bibfnamefont {A.~J.}\ \bibnamefont
  {Mathijssen}}, \bibinfo {author} {\bibfnamefont {C.}~\bibnamefont {Hoell}},
  \bibinfo {author} {\bibfnamefont {S.}~\bibnamefont {Goh}}, \bibinfo {author}
  {\bibfnamefont {J.}~\bibnamefont {Blawzdziewicz}}, \bibinfo {author}
  {\bibfnamefont {A.~M.}\ \bibnamefont {Menzel}}, \ and\ \bibinfo {author}
  {\bibfnamefont {H.}~\bibnamefont {L{\"o}wen}},\ }\href@noop {} {\bibfield
  {journal} {\bibinfo  {journal} {J. Phys.: Condens. Matter}\ }\textbf
  {\bibinfo {volume} {30}},\ \bibinfo {pages} {254004} (\bibinfo {year}
  {2018}{\natexlab{b}})}\BibitemShut {NoStop}%
\bibitem [{\citenamefont {Underhill}, \citenamefont {Hernandez-Ortiz},\ and\
  \citenamefont {Graham}(2008)}]{underhill2008diffusion}%
  \BibitemOpen
  \bibfield  {author} {\bibinfo {author} {\bibfnamefont {P.~T.}\ \bibnamefont
  {Underhill}}, \bibinfo {author} {\bibfnamefont {J.~P.}\ \bibnamefont
  {Hernandez-Ortiz}}, \ and\ \bibinfo {author} {\bibfnamefont {M.~D.}\
  \bibnamefont {Graham}},\ }\href@noop {} {\bibfield  {journal} {\bibinfo
  {journal} {Phys. Rev. Lett.}\ }\textbf {\bibinfo {volume} {100}},\ \bibinfo
  {pages} {248101} (\bibinfo {year} {2008})}\BibitemShut {NoStop}%
\bibitem [{\citenamefont {Baskaran}\ and\ \citenamefont
  {Marchetti}(2009)}]{baskaran2009statistical}%
  \BibitemOpen
  \bibfield  {author} {\bibinfo {author} {\bibfnamefont {A.}~\bibnamefont
  {Baskaran}}\ and\ \bibinfo {author} {\bibfnamefont {M.~C.}\ \bibnamefont
  {Marchetti}},\ }\href@noop {} {\bibfield  {journal} {\bibinfo  {journal}
  {Proc. Natl. Acad. Sci. U.S.A.}\ }\textbf {\bibinfo {volume} {106}},\
  \bibinfo {pages} {15567} (\bibinfo {year} {2009})}\BibitemShut {NoStop}%
\bibitem [{\citenamefont {Vicsek}\ \emph {et~al.}(1995)\citenamefont {Vicsek},
  \citenamefont {Czir{\'o}k}, \citenamefont {Ben-Jacob}, \citenamefont
  {Cohen},\ and\ \citenamefont {Shochet}}]{vicsek1995novel}%
  \BibitemOpen
  \bibfield  {author} {\bibinfo {author} {\bibfnamefont {T.}~\bibnamefont
  {Vicsek}}, \bibinfo {author} {\bibfnamefont {A.}~\bibnamefont {Czir{\'o}k}},
  \bibinfo {author} {\bibfnamefont {E.}~\bibnamefont {Ben-Jacob}}, \bibinfo
  {author} {\bibfnamefont {I.}~\bibnamefont {Cohen}}, \ and\ \bibinfo {author}
  {\bibfnamefont {O.}~\bibnamefont {Shochet}},\ }\href@noop {} {\bibfield
  {journal} {\bibinfo  {journal} {Phys. Rev. Lett.}\ }\textbf {\bibinfo
  {volume} {75}},\ \bibinfo {pages} {1226} (\bibinfo {year}
  {1995})}\BibitemShut {NoStop}%
\bibitem [{\citenamefont {Toner}\ and\ \citenamefont
  {Tu}(1995)}]{toner1995long}%
  \BibitemOpen
  \bibfield  {author} {\bibinfo {author} {\bibfnamefont {J.}~\bibnamefont
  {Toner}}\ and\ \bibinfo {author} {\bibfnamefont {Y.}~\bibnamefont {Tu}},\
  }\href@noop {} {\bibfield  {journal} {\bibinfo  {journal} {Phys. Rev. Lett.}\
  }\textbf {\bibinfo {volume} {75}},\ \bibinfo {pages} {4326} (\bibinfo {year}
  {1995})}\BibitemShut {NoStop}%
\bibitem [{\citenamefont {Farrell}\ \emph {et~al.}(2012)\citenamefont
  {Farrell}, \citenamefont {Marchetti}, \citenamefont {Marenduzzo},\ and\
  \citenamefont {Tailleur}}]{farrell2012pattern}%
  \BibitemOpen
  \bibfield  {author} {\bibinfo {author} {\bibfnamefont {F.~D.~C.}\
  \bibnamefont {Farrell}}, \bibinfo {author} {\bibfnamefont {M.~C.}\
  \bibnamefont {Marchetti}}, \bibinfo {author} {\bibfnamefont {D.}~\bibnamefont
  {Marenduzzo}}, \ and\ \bibinfo {author} {\bibfnamefont {J.}~\bibnamefont
  {Tailleur}},\ }\href@noop {} {\bibfield  {journal} {\bibinfo  {journal}
  {Phys. Rev. Lett.}\ }\textbf {\bibinfo {volume} {108}},\ \bibinfo {pages}
  {248101} (\bibinfo {year} {2012})}\BibitemShut {NoStop}%
\bibitem [{\citenamefont {Vicsek}\ and\ \citenamefont
  {Zafeiris}(2012)}]{vicsek2012collective}%
  \BibitemOpen
  \bibfield  {author} {\bibinfo {author} {\bibfnamefont {T.}~\bibnamefont
  {Vicsek}}\ and\ \bibinfo {author} {\bibfnamefont {A.}~\bibnamefont
  {Zafeiris}},\ }\href@noop {} {\bibfield  {journal} {\bibinfo  {journal}
  {Phys. Rep.}\ }\textbf {\bibinfo {volume} {517}},\ \bibinfo {pages} {71}
  (\bibinfo {year} {2012})}\BibitemShut {NoStop}%
\bibitem [{\citenamefont {Liebchen}\ and\ \citenamefont
  {Levis}(2017)}]{liebchen2017collective}%
  \BibitemOpen
  \bibfield  {author} {\bibinfo {author} {\bibfnamefont {B.}~\bibnamefont
  {Liebchen}}\ and\ \bibinfo {author} {\bibfnamefont {D.}~\bibnamefont
  {Levis}},\ }\href@noop {} {\bibfield  {journal} {\bibinfo  {journal} {Phys.
  Rev. Lett.}\ }\textbf {\bibinfo {volume} {119}},\ \bibinfo {pages} {058002}
  (\bibinfo {year} {2017})}\BibitemShut {NoStop}%
\bibitem [{\citenamefont {Levis}\ and\ \citenamefont
  {Liebchen}(2018)}]{levis2018micro}%
  \BibitemOpen
  \bibfield  {author} {\bibinfo {author} {\bibfnamefont {D.}~\bibnamefont
  {Levis}}\ and\ \bibinfo {author} {\bibfnamefont {B.}~\bibnamefont
  {Liebchen}},\ }\href@noop {} {\bibfield  {journal} {\bibinfo  {journal} {J.
  Phys.: Condens. Matter}\ }\textbf {\bibinfo {volume} {30}},\ \bibinfo {pages}
  {084001} (\bibinfo {year} {2018})}\BibitemShut {NoStop}%
\bibitem [{\citenamefont {Ginelli}\ \emph {et~al.}(2010)\citenamefont
  {Ginelli}, \citenamefont {Peruani}, \citenamefont {B{\"a}r},\ and\
  \citenamefont {Chat{\'e}}}]{ginelli2010large}%
  \BibitemOpen
  \bibfield  {author} {\bibinfo {author} {\bibfnamefont {F.}~\bibnamefont
  {Ginelli}}, \bibinfo {author} {\bibfnamefont {F.}~\bibnamefont {Peruani}},
  \bibinfo {author} {\bibfnamefont {M.}~\bibnamefont {B{\"a}r}}, \ and\
  \bibinfo {author} {\bibfnamefont {H.}~\bibnamefont {Chat{\'e}}},\ }\href@noop
  {} {\bibfield  {journal} {\bibinfo  {journal} {Phys. Rev. Lett.}\ }\textbf
  {\bibinfo {volume} {104}},\ \bibinfo {pages} {184502} (\bibinfo {year}
  {2010})}\BibitemShut {NoStop}%
\bibitem [{\citenamefont {Alarc{\'o}n}\ and\ \citenamefont
  {Pagonabarraga}(2013)}]{alarcon2013spontaneous}%
  \BibitemOpen
  \bibfield  {author} {\bibinfo {author} {\bibfnamefont {F.}~\bibnamefont
  {Alarc{\'o}n}}\ and\ \bibinfo {author} {\bibfnamefont {I.}~\bibnamefont
  {Pagonabarraga}},\ }\href@noop {} {\bibfield  {journal} {\bibinfo  {journal}
  {J. Mol. Liq.}\ }\textbf {\bibinfo {volume} {185}},\ \bibinfo {pages} {56}
  (\bibinfo {year} {2013})}\BibitemShut {NoStop}%
\bibitem [{\citenamefont {Pessot}, \citenamefont {{L\"owen}},\ and\
  \citenamefont {Menzel}(2018)}]{pessot2018}%
  \BibitemOpen
  \bibfield  {author} {\bibinfo {author} {\bibfnamefont {G.}~\bibnamefont
  {Pessot}}, \bibinfo {author} {\bibfnamefont {H.}~\bibnamefont {{L\"owen}}}, \
  and\ \bibinfo {author} {\bibfnamefont {A.~M.}\ \bibnamefont {Menzel}},\
  }\href@noop {} {\bibfield  {journal} {\bibinfo  {journal} {Mol. Phys.}\
  }\textbf {\bibinfo {volume} {116}},\ \bibinfo {pages} {3401} (\bibinfo {year}
  {2018})}\BibitemShut {NoStop}%
\bibitem [{\citenamefont {Menzel}\ \emph {et~al.}(2016)\citenamefont {Menzel},
  \citenamefont {Saha}, \citenamefont {Hoell},\ and\ \citenamefont
  {L{\"o}wen}}]{menzel2016dynamical}%
  \BibitemOpen
  \bibfield  {author} {\bibinfo {author} {\bibfnamefont {A.~M.}\ \bibnamefont
  {Menzel}}, \bibinfo {author} {\bibfnamefont {A.}~\bibnamefont {Saha}},
  \bibinfo {author} {\bibfnamefont {C.}~\bibnamefont {Hoell}}, \ and\ \bibinfo
  {author} {\bibfnamefont {H.}~\bibnamefont {L{\"o}wen}},\ }\href@noop {}
  {\bibfield  {journal} {\bibinfo  {journal} {J. Chem. Phys.}\ }\textbf
  {\bibinfo {volume} {144}},\ \bibinfo {pages} {024115} (\bibinfo {year}
  {2016})}\BibitemShut {NoStop}%
\bibitem [{\citenamefont {Hoell}, \citenamefont {L\"owen},\ and\ \citenamefont
  {Menzel}(2017)}]{hoell2017dynamical}%
  \BibitemOpen
  \bibfield  {author} {\bibinfo {author} {\bibfnamefont {C.}~\bibnamefont
  {Hoell}}, \bibinfo {author} {\bibfnamefont {H.}~\bibnamefont {L\"owen}}, \
  and\ \bibinfo {author} {\bibfnamefont {A.~M.}\ \bibnamefont {Menzel}},\
  }\href@noop {} {\bibfield  {journal} {\bibinfo  {journal} {New J. Phys.}\
  }\textbf {\bibinfo {volume} {19}},\ \bibinfo {pages} {125004} (\bibinfo
  {year} {2017})}\BibitemShut {NoStop}%
\bibitem [{\citenamefont {Beenakker}\ and\ \citenamefont
  {Mazur}(1983)}]{beenakker1983self}%
  \BibitemOpen
  \bibfield  {author} {\bibinfo {author} {\bibfnamefont {C.~W.~J.}\
  \bibnamefont {Beenakker}}\ and\ \bibinfo {author} {\bibfnamefont
  {P.}~\bibnamefont {Mazur}},\ }\href@noop {} {\bibfield  {journal} {\bibinfo
  {journal} {Physica A}\ }\textbf {\bibinfo {volume} {120}},\ \bibinfo {pages}
  {388} (\bibinfo {year} {1983})}\BibitemShut {NoStop}%
\bibitem [{\citenamefont {Beenakker}\ and\ \citenamefont
  {Mazur}(1984)}]{beenakker1984diffusion}%
  \BibitemOpen
  \bibfield  {author} {\bibinfo {author} {\bibfnamefont {C.~W.~J.}\
  \bibnamefont {Beenakker}}\ and\ \bibinfo {author} {\bibfnamefont
  {P.}~\bibnamefont {Mazur}},\ }\href@noop {} {\bibfield  {journal} {\bibinfo
  {journal} {Physica A}\ }\textbf {\bibinfo {volume} {126}},\ \bibinfo {pages}
  {349} (\bibinfo {year} {1984})}\BibitemShut {NoStop}%
\bibitem [{\citenamefont {N{\"a}gele}(1996)}]{nagele1996dynamics}%
  \BibitemOpen
  \bibfield  {author} {\bibinfo {author} {\bibfnamefont {G.}~\bibnamefont
  {N{\"a}gele}},\ }\href@noop {} {\bibfield  {journal} {\bibinfo  {journal}
  {Phys. Rep.}\ }\textbf {\bibinfo {volume} {272}},\ \bibinfo {pages} {215}
  (\bibinfo {year} {1996})}\BibitemShut {NoStop}%
\bibitem [{\citenamefont {N{\"a}gele}\ and\ \citenamefont
  {Baur}(1997)}]{nagele1997long}%
  \BibitemOpen
  \bibfield  {author} {\bibinfo {author} {\bibfnamefont {G.}~\bibnamefont
  {N{\"a}gele}}\ and\ \bibinfo {author} {\bibfnamefont {P.}~\bibnamefont
  {Baur}},\ }\href@noop {} {\bibfield  {journal} {\bibinfo  {journal} {Physica
  A}\ }\textbf {\bibinfo {volume} {245}},\ \bibinfo {pages} {297} (\bibinfo
  {year} {1997})}\BibitemShut {NoStop}%
\bibitem [{\citenamefont {Banchio}\ \emph {et~al.}(2006)\citenamefont
  {Banchio}, \citenamefont {Gapinski}, \citenamefont {Patkowski}, \citenamefont
  {H{\"a}u{\ss}ler}, \citenamefont {Fluerasu}, \citenamefont {Sacanna},
  \citenamefont {Holmqvist}, \citenamefont {Meier}, \citenamefont {Lettinga},\
  and\ \citenamefont {N{\"a}gele}}]{banchio2006many}%
  \BibitemOpen
  \bibfield  {author} {\bibinfo {author} {\bibfnamefont {A.~J.}\ \bibnamefont
  {Banchio}}, \bibinfo {author} {\bibfnamefont {J.}~\bibnamefont {Gapinski}},
  \bibinfo {author} {\bibfnamefont {A.}~\bibnamefont {Patkowski}}, \bibinfo
  {author} {\bibfnamefont {W.}~\bibnamefont {H{\"a}u{\ss}ler}}, \bibinfo
  {author} {\bibfnamefont {A.}~\bibnamefont {Fluerasu}}, \bibinfo {author}
  {\bibfnamefont {S.}~\bibnamefont {Sacanna}}, \bibinfo {author} {\bibfnamefont
  {P.}~\bibnamefont {Holmqvist}}, \bibinfo {author} {\bibfnamefont
  {G.}~\bibnamefont {Meier}}, \bibinfo {author} {\bibfnamefont {M.~P.}\
  \bibnamefont {Lettinga}}, \ and\ \bibinfo {author} {\bibfnamefont
  {G.}~\bibnamefont {N{\"a}gele}},\ }\href@noop {} {\bibfield  {journal}
  {\bibinfo  {journal} {Phys. Rev. Lett.}\ }\textbf {\bibinfo {volume} {96}},\
  \bibinfo {pages} {138303} (\bibinfo {year} {2006})}\BibitemShut {NoStop}%
\bibitem [{\citenamefont {Rotne}\ and\ \citenamefont
  {Prager}(1969)}]{Rotne_1969_JCP}%
  \BibitemOpen
  \bibfield  {author} {\bibinfo {author} {\bibfnamefont {J.}~\bibnamefont
  {Rotne}}\ and\ \bibinfo {author} {\bibfnamefont {S.}~\bibnamefont {Prager}},\
  }\href@noop {} {\bibfield  {journal} {\bibinfo  {journal} {J. Chem. Phys.}\
  }\textbf {\bibinfo {volume} {50}},\ \bibinfo {pages} {4831} (\bibinfo {year}
  {1969})}\BibitemShut {NoStop}%
\bibitem [{\citenamefont {Reichert}\ and\ \citenamefont
  {Stark}(2004)}]{reichert2004hydrodynamic}%
  \BibitemOpen
  \bibfield  {author} {\bibinfo {author} {\bibfnamefont {M.}~\bibnamefont
  {Reichert}}\ and\ \bibinfo {author} {\bibfnamefont {H.}~\bibnamefont
  {Stark}},\ }\href@noop {} {\bibfield  {journal} {\bibinfo  {journal} {Phys.
  Rev. E}\ }\textbf {\bibinfo {volume} {69}},\ \bibinfo {pages} {031407}
  (\bibinfo {year} {2004})}\BibitemShut {NoStop}%
\bibitem [{\citenamefont {Adhyapak}\ and\ \citenamefont
  {Jabbari-Farouji}(2017)}]{adhyapak2017flow}%
  \BibitemOpen
  \bibfield  {author} {\bibinfo {author} {\bibfnamefont {T.~C.}\ \bibnamefont
  {Adhyapak}}\ and\ \bibinfo {author} {\bibfnamefont {S.}~\bibnamefont
  {Jabbari-Farouji}},\ }\href@noop {} {\bibfield  {journal} {\bibinfo
  {journal} {Phys. Rev. E}\ }\textbf {\bibinfo {volume} {96}},\ \bibinfo
  {pages} {052608} (\bibinfo {year} {2017})}\BibitemShut {NoStop}%
\bibitem [{Note1()}]{Note1}%
  \BibitemOpen
  \bibinfo {note} {Strictly speaking, only the lowest-order terms in an
  expansion around $a/L=0$ are included for $j \not =i$ in Eqs.~(\ref
  {mu_tt_pm}) and (\ref {mu_rt_pm}). We neglect higher-order corrections\cite
  {Kim_Karrila, adhyapak2017flow} in favor of simplicity and more traceable
  analytical expressions.}\BibitemShut {Stop}%
\bibitem [{\citenamefont {Hansen}\ and\ \citenamefont
  {McDonald}(1990)}]{hansen1990theory}%
  \BibitemOpen
  \bibfield  {author} {\bibinfo {author} {\bibfnamefont {J.-P.}\ \bibnamefont
  {Hansen}}\ and\ \bibinfo {author} {\bibfnamefont {I.~R.}\ \bibnamefont
  {McDonald}},\ }\href@noop {} {\emph {\bibinfo {title} {Theory of Simple
  Liquids}}}\ (\bibinfo  {publisher} {Elsevier, Amsterdam},\ \bibinfo {year}
  {1990})\BibitemShut {NoStop}%
\bibitem [{\citenamefont {Evans}(1979)}]{evans1979nature}%
  \BibitemOpen
  \bibfield  {author} {\bibinfo {author} {\bibfnamefont {R.}~\bibnamefont
  {Evans}},\ }\href@noop {} {\bibfield  {journal} {\bibinfo  {journal} {Adv.
  Phys.}\ }\textbf {\bibinfo {volume} {28}},\ \bibinfo {pages} {143} (\bibinfo
  {year} {1979})}\BibitemShut {NoStop}%
\bibitem [{\citenamefont {Evans}(1992)}]{evans1992density}%
  \BibitemOpen
  \bibfield  {author} {\bibinfo {author} {\bibfnamefont {R.}~\bibnamefont
  {Evans}},\ }in\ \href@noop {} {\emph {\bibinfo {booktitle} {Fundamentals of
  Inhomogeneous Fluids}}},\ \bibinfo {editor} {edited by\ \bibinfo {editor}
  {\bibfnamefont {D.}~\bibnamefont {Henderson}}}\ (\bibinfo  {publisher}
  {Marcel Dekker, New York},\ \bibinfo {year} {1992})\ pp.\ \bibinfo {pages}
  {85--176}\BibitemShut {NoStop}%
\bibitem [{\citenamefont {Marconi}\ and\ \citenamefont
  {Tarazona}(1999)}]{marconi1999dynamic}%
  \BibitemOpen
  \bibfield  {author} {\bibinfo {author} {\bibfnamefont {U.~M.~B.}\
  \bibnamefont {Marconi}}\ and\ \bibinfo {author} {\bibfnamefont
  {P.}~\bibnamefont {Tarazona}},\ }\href@noop {} {\bibfield  {journal}
  {\bibinfo  {journal} {J. Chem. Phys.}\ }\textbf {\bibinfo {volume} {110}},\
  \bibinfo {pages} {8032} (\bibinfo {year} {1999})}\BibitemShut {NoStop}%
\bibitem [{\citenamefont {Marconi}\ and\ \citenamefont
  {Tarazona}(2000)}]{marconi2000dynamic}%
  \BibitemOpen
  \bibfield  {author} {\bibinfo {author} {\bibfnamefont {U.~M.~B.}\
  \bibnamefont {Marconi}}\ and\ \bibinfo {author} {\bibfnamefont
  {P.}~\bibnamefont {Tarazona}},\ }\href@noop {} {\bibfield  {journal}
  {\bibinfo  {journal} {J. Phys.: Condens. Matter}\ }\textbf {\bibinfo {volume}
  {12}},\ \bibinfo {pages} {A413} (\bibinfo {year} {2000})}\BibitemShut
  {NoStop}%
\bibitem [{\citenamefont {Archer}\ and\ \citenamefont
  {Evans}(2004)}]{archer2004dynamical}%
  \BibitemOpen
  \bibfield  {author} {\bibinfo {author} {\bibfnamefont {A.~J.}\ \bibnamefont
  {Archer}}\ and\ \bibinfo {author} {\bibfnamefont {R.}~\bibnamefont {Evans}},\
  }\href@noop {} {\bibfield  {journal} {\bibinfo  {journal} {J. Chem. Phys.}\
  }\textbf {\bibinfo {volume} {121}},\ \bibinfo {pages} {4246} (\bibinfo {year}
  {2004})}\BibitemShut {NoStop}%
\bibitem [{\citenamefont {Chan}\ and\ \citenamefont
  {Finken}(2005)}]{chan2005time}%
  \BibitemOpen
  \bibfield  {author} {\bibinfo {author} {\bibfnamefont {G.~K.-L.}\
  \bibnamefont {Chan}}\ and\ \bibinfo {author} {\bibfnamefont {R.}~\bibnamefont
  {Finken}},\ }\href@noop {} {\bibfield  {journal} {\bibinfo  {journal} {Phys.
  Rev. Lett.}\ }\textbf {\bibinfo {volume} {94}},\ \bibinfo {pages} {183001}
  (\bibinfo {year} {2005})}\BibitemShut {NoStop}%
\bibitem [{\citenamefont {Espa{\~n}ol}\ and\ \citenamefont
  {L{\"o}wen}(2009)}]{espanol2009derivation}%
  \BibitemOpen
  \bibfield  {author} {\bibinfo {author} {\bibfnamefont {P.}~\bibnamefont
  {Espa{\~n}ol}}\ and\ \bibinfo {author} {\bibfnamefont {H.}~\bibnamefont
  {L{\"o}wen}},\ }\href@noop {} {\bibfield  {journal} {\bibinfo  {journal} {J.
  Chem. Phys.}\ }\textbf {\bibinfo {volume} {131}},\ \bibinfo {pages} {244101}
  (\bibinfo {year} {2009})}\BibitemShut {NoStop}%
\bibitem [{\citenamefont {Evans}(2010)}]{evans2010density}%
  \BibitemOpen
  \bibfield  {author} {\bibinfo {author} {\bibfnamefont {R.}~\bibnamefont
  {Evans}},\ }in\ \href@noop {} {\emph {\bibinfo {booktitle} {Lecture Notes 3rd
  Warsaw School of Statistical Physics}}},\ \bibinfo {editor} {edited by\
  \bibinfo {editor} {\bibfnamefont {B.}~\bibnamefont {Cichocki}}, \bibinfo
  {editor} {\bibfnamefont {M.}~\bibnamefont {Napi\'{o}rkowski}}, \ and\
  \bibinfo {editor} {\bibfnamefont {J.}~\bibnamefont {Piasecki}}}\ (\bibinfo
  {publisher} {Warsaw University Press, Warsaw},\ \bibinfo {year} {2010})\ pp.\
  \bibinfo {pages} {43--85}\BibitemShut {NoStop}%
\bibitem [{\citenamefont {L{\"o}wen}(2010)}]{lowen2010density}%
  \BibitemOpen
  \bibfield  {author} {\bibinfo {author} {\bibfnamefont {H.}~\bibnamefont
  {L{\"o}wen}},\ }in\ \href@noop {} {\emph {\bibinfo {booktitle} {Lecture Notes
  3rd Warsaw School of Statistical Physics}}},\ \bibinfo {editor} {edited by\
  \bibinfo {editor} {\bibfnamefont {B.}~\bibnamefont {Cichocki}}, \bibinfo
  {editor} {\bibfnamefont {M.}~\bibnamefont {Napi\'{o}rkowski}}, \ and\
  \bibinfo {editor} {\bibfnamefont {J.}~\bibnamefont {Piasecki}}}\ (\bibinfo
  {publisher} {Warsaw University Press, Warsaw},\ \bibinfo {year} {2010})\ pp.\
  \bibinfo {pages} {87--121}\BibitemShut {NoStop}%
\bibitem [{\citenamefont {Wittkowski}\ and\ \citenamefont
  {L{\"o}wen}(2011)}]{wittkowski2011dynamical}%
  \BibitemOpen
  \bibfield  {author} {\bibinfo {author} {\bibfnamefont {R.}~\bibnamefont
  {Wittkowski}}\ and\ \bibinfo {author} {\bibfnamefont {H.}~\bibnamefont
  {L{\"o}wen}},\ }\href@noop {} {\bibfield  {journal} {\bibinfo  {journal}
  {Mol. Phys.}\ }\textbf {\bibinfo {volume} {109}},\ \bibinfo {pages} {2935}
  (\bibinfo {year} {2011})}\BibitemShut {NoStop}%
\bibitem [{\citenamefont {Lugo-Fr{\'\i}as}\ and\ \citenamefont
  {Klapp}(2016)}]{lugo2016binary}%
  \BibitemOpen
  \bibfield  {author} {\bibinfo {author} {\bibfnamefont {R.}~\bibnamefont
  {Lugo-Fr{\'\i}as}}\ and\ \bibinfo {author} {\bibfnamefont {S.~H.}\
  \bibnamefont {Klapp}},\ }\href@noop {} {\bibfield  {journal} {\bibinfo
  {journal} {J. Phys.: Condens. Matter}\ }\textbf {\bibinfo {volume} {28}},\
  \bibinfo {pages} {244022} (\bibinfo {year} {2016})}\BibitemShut {NoStop}%
\bibitem [{\citenamefont {Gray}\ and\ \citenamefont
  {Gubbins}(1984)}]{Gray_Gubbins}%
  \BibitemOpen
  \bibfield  {author} {\bibinfo {author} {\bibfnamefont {C.~G.}\ \bibnamefont
  {Gray}}\ and\ \bibinfo {author} {\bibfnamefont {K.~E.}\ \bibnamefont
  {Gubbins}},\ }\href@noop {} {\emph {\bibinfo {title} {Theory of Molecular
  Fluids: I: Fundamentals}}},\ (International Series of Monographs on
  Chemistry)\ (\bibinfo  {publisher} {Oxford University Press, Oxford},\
  \bibinfo {year} {1984})\BibitemShut {NoStop}%
\bibitem [{\citenamefont {Percus}(1962)}]{percus1962approximation}%
  \BibitemOpen
  \bibfield  {author} {\bibinfo {author} {\bibfnamefont {J.~K.}\ \bibnamefont
  {Percus}},\ }\href@noop {} {\bibfield  {journal} {\bibinfo  {journal} {Phys.
  Rev. Lett.}\ }\textbf {\bibinfo {volume} {8}},\ \bibinfo {pages} {462}
  (\bibinfo {year} {1962})}\BibitemShut {NoStop}%
\bibitem [{\citenamefont {Archer}, \citenamefont {Chacko},\ and\ \citenamefont
  {Evans}(2017)}]{archer2017standard}%
  \BibitemOpen
  \bibfield  {author} {\bibinfo {author} {\bibfnamefont {A.~J.}\ \bibnamefont
  {Archer}}, \bibinfo {author} {\bibfnamefont {B.}~\bibnamefont {Chacko}}, \
  and\ \bibinfo {author} {\bibfnamefont {R.}~\bibnamefont {Evans}},\
  }\href@noop {} {\bibfield  {journal} {\bibinfo  {journal} {J. Chem. Phys.}\
  }\textbf {\bibinfo {volume} {147}},\ \bibinfo {pages} {034501} (\bibinfo
  {year} {2017})}\BibitemShut {NoStop}%
\bibitem [{\citenamefont {Hopkins}\ \emph {et~al.}(2010)\citenamefont
  {Hopkins}, \citenamefont {Fortini}, \citenamefont {Archer},\ and\
  \citenamefont {Schmidt}}]{hopkins2010van}%
  \BibitemOpen
  \bibfield  {author} {\bibinfo {author} {\bibfnamefont {P.}~\bibnamefont
  {Hopkins}}, \bibinfo {author} {\bibfnamefont {A.}~\bibnamefont {Fortini}},
  \bibinfo {author} {\bibfnamefont {A.~J.}\ \bibnamefont {Archer}}, \ and\
  \bibinfo {author} {\bibfnamefont {M.}~\bibnamefont {Schmidt}},\ }\href@noop
  {} {\bibfield  {journal} {\bibinfo  {journal} {J. Chem. Phys.}\ }\textbf
  {\bibinfo {volume} {133}},\ \bibinfo {pages} {224505} (\bibinfo {year}
  {2010})}\BibitemShut {NoStop}%
\bibitem [{\citenamefont {Brader}\ and\ \citenamefont
  {Schmidt}(2015)}]{brader2015power}%
  \BibitemOpen
  \bibfield  {author} {\bibinfo {author} {\bibfnamefont {J.~M.}\ \bibnamefont
  {Brader}}\ and\ \bibinfo {author} {\bibfnamefont {M.}~\bibnamefont
  {Schmidt}},\ }\href@noop {} {\bibfield  {journal} {\bibinfo  {journal} {J.
  Phys.: Condens. Matter}\ }\textbf {\bibinfo {volume} {27}},\ \bibinfo {pages}
  {194106} (\bibinfo {year} {2015})}\BibitemShut {NoStop}%
\bibitem [{\citenamefont {Rein}\ and\ \citenamefont
  {Speck}(2016)}]{rein2016applicability}%
  \BibitemOpen
  \bibfield  {author} {\bibinfo {author} {\bibfnamefont {M.}~\bibnamefont
  {Rein}}\ and\ \bibinfo {author} {\bibfnamefont {T.}~\bibnamefont {Speck}},\
  }\href@noop {} {\bibfield  {journal} {\bibinfo  {journal} {Eur. Phys. J. E}\
  }\textbf {\bibinfo {volume} {39}},\ \bibinfo {pages} {84} (\bibinfo {year}
  {2016})}\BibitemShut {NoStop}%
\bibitem [{\citenamefont {H{\"a}rtel}, \citenamefont {Richard},\ and\
  \citenamefont {Speck}(2018)}]{hartel2018three}%
  \BibitemOpen
  \bibfield  {author} {\bibinfo {author} {\bibfnamefont {A.}~\bibnamefont
  {H{\"a}rtel}}, \bibinfo {author} {\bibfnamefont {D.}~\bibnamefont {Richard}},
  \ and\ \bibinfo {author} {\bibfnamefont {T.}~\bibnamefont {Speck}},\
  }\href@noop {} {\bibfield  {journal} {\bibinfo  {journal} {Phys. Rev. E}\
  }\textbf {\bibinfo {volume} {97}},\ \bibinfo {pages} {012606} (\bibinfo
  {year} {2018})}\BibitemShut {NoStop}%
\bibitem [{\citenamefont {Stenhammar}\ \emph {et~al.}(2013)\citenamefont
  {Stenhammar}, \citenamefont {Tiribocchi}, \citenamefont {Allen},
  \citenamefont {Marenduzzo},\ and\ \citenamefont
  {Cates}}]{stenhammar2013continuum}%
  \BibitemOpen
  \bibfield  {author} {\bibinfo {author} {\bibfnamefont {J.}~\bibnamefont
  {Stenhammar}}, \bibinfo {author} {\bibfnamefont {A.}~\bibnamefont
  {Tiribocchi}}, \bibinfo {author} {\bibfnamefont {R.~J.}\ \bibnamefont
  {Allen}}, \bibinfo {author} {\bibfnamefont {D.}~\bibnamefont {Marenduzzo}}, \
  and\ \bibinfo {author} {\bibfnamefont {M.~E.}\ \bibnamefont {Cates}},\
  }\href@noop {} {\bibfield  {journal} {\bibinfo  {journal} {Phys. Rev. Lett.}\
  }\textbf {\bibinfo {volume} {111}},\ \bibinfo {pages} {145702} (\bibinfo
  {year} {2013})}\BibitemShut {NoStop}%
\bibitem [{\citenamefont {Wensink}\ and\ \citenamefont
  {L{\"o}wen}(2008)}]{wensink2008aggregation}%
  \BibitemOpen
  \bibfield  {author} {\bibinfo {author} {\bibfnamefont {H.~H.}\ \bibnamefont
  {Wensink}}\ and\ \bibinfo {author} {\bibfnamefont {H.}~\bibnamefont
  {L{\"o}wen}},\ }\href@noop {} {\bibfield  {journal} {\bibinfo  {journal}
  {Phys. Rev. E}\ }\textbf {\bibinfo {volume} {78}},\ \bibinfo {pages} {031409}
  (\bibinfo {year} {2008})}\BibitemShut {NoStop}%
\bibitem [{\citenamefont {Mladek}\ \emph {et~al.}(2006)\citenamefont {Mladek},
  \citenamefont {Gottwald}, \citenamefont {Kahl}, \citenamefont {Neumann},\
  and\ \citenamefont {Likos}}]{mladek2006formation}%
  \BibitemOpen
  \bibfield  {author} {\bibinfo {author} {\bibfnamefont {B.~M.}\ \bibnamefont
  {Mladek}}, \bibinfo {author} {\bibfnamefont {D.}~\bibnamefont {Gottwald}},
  \bibinfo {author} {\bibfnamefont {G.}~\bibnamefont {Kahl}}, \bibinfo {author}
  {\bibfnamefont {M.}~\bibnamefont {Neumann}}, \ and\ \bibinfo {author}
  {\bibfnamefont {C.~N.}\ \bibnamefont {Likos}},\ }\href@noop {} {\bibfield
  {journal} {\bibinfo  {journal} {Phys. Rev. Lett.}\ }\textbf {\bibinfo
  {volume} {96}},\ \bibinfo {pages} {045701} (\bibinfo {year}
  {2006})}\BibitemShut {NoStop}%
\bibitem [{\citenamefont {Archer}\ \emph {et~al.}(2014)\citenamefont {Archer},
  \citenamefont {Walters}, \citenamefont {Thiele},\ and\ \citenamefont
  {Knobloch}}]{archer2014solidification}%
  \BibitemOpen
  \bibfield  {author} {\bibinfo {author} {\bibfnamefont {A.~J.}\ \bibnamefont
  {Archer}}, \bibinfo {author} {\bibfnamefont {M.}~\bibnamefont {Walters}},
  \bibinfo {author} {\bibfnamefont {U.}~\bibnamefont {Thiele}}, \ and\ \bibinfo
  {author} {\bibfnamefont {E.}~\bibnamefont {Knobloch}},\ }\href@noop {}
  {\bibfield  {journal} {\bibinfo  {journal} {Phys. Rev. E}\ }\textbf {\bibinfo
  {volume} {90}},\ \bibinfo {pages} {042404} (\bibinfo {year}
  {2014})}\BibitemShut {NoStop}%
\bibitem [{\citenamefont {Guyer}, \citenamefont {Wheeler},\ and\ \citenamefont
  {Warren}(2009)}]{Guyer_2009_CiSE}%
  \BibitemOpen
  \bibfield  {author} {\bibinfo {author} {\bibfnamefont {J.~E.}\ \bibnamefont
  {Guyer}}, \bibinfo {author} {\bibfnamefont {D.}~\bibnamefont {Wheeler}}, \
  and\ \bibinfo {author} {\bibfnamefont {J.~A.}\ \bibnamefont {Warren}},\
  }\href@noop {} {\bibfield  {journal} {\bibinfo  {journal} {Comput. Sci.
  Eng.}\ }\textbf {\bibinfo {volume} {11}},\ \bibinfo {pages} {6} (\bibinfo
  {year} {2009})}\BibitemShut {NoStop}%
\bibitem [{\citenamefont {Stenhammar}\ \emph {et~al.}(2017)\citenamefont
  {Stenhammar}, \citenamefont {Nardini}, \citenamefont {Nash}, \citenamefont
  {Marenduzzo},\ and\ \citenamefont {Morozov}}]{stenhammar2017role}%
  \BibitemOpen
  \bibfield  {author} {\bibinfo {author} {\bibfnamefont {J.}~\bibnamefont
  {Stenhammar}}, \bibinfo {author} {\bibfnamefont {C.}~\bibnamefont {Nardini}},
  \bibinfo {author} {\bibfnamefont {R.~W.}\ \bibnamefont {Nash}}, \bibinfo
  {author} {\bibfnamefont {D.}~\bibnamefont {Marenduzzo}}, \ and\ \bibinfo
  {author} {\bibfnamefont {A.}~\bibnamefont {Morozov}},\ }\href@noop {}
  {\bibfield  {journal} {\bibinfo  {journal} {Phys. Rev. Lett.}\ }\textbf
  {\bibinfo {volume} {119}},\ \bibinfo {pages} {028005} (\bibinfo {year}
  {2017})}\BibitemShut {NoStop}%
\bibitem [{\citenamefont {de~Macedo~Biniossek}\ \emph
  {et~al.}(2018)\citenamefont {de~Macedo~Biniossek}, \citenamefont {L{\"o}wen},
  \citenamefont {Voigtmann},\ and\ \citenamefont
  {Smallenburg}}]{nadine2018static}%
  \BibitemOpen
  \bibfield  {author} {\bibinfo {author} {\bibfnamefont {N.}~\bibnamefont
  {de~Macedo~Biniossek}}, \bibinfo {author} {\bibfnamefont {H.}~\bibnamefont
  {L{\"o}wen}}, \bibinfo {author} {\bibfnamefont {T.}~\bibnamefont
  {Voigtmann}}, \ and\ \bibinfo {author} {\bibfnamefont {F.}~\bibnamefont
  {Smallenburg}},\ }\href@noop {} {\bibfield  {journal} {\bibinfo  {journal}
  {J. Phys.: Condens. Matter}\ }\textbf {\bibinfo {volume} {30}},\ \bibinfo
  {pages} {074001} (\bibinfo {year} {2018})}\BibitemShut {NoStop}%
\bibitem [{\citenamefont {Heidenreich}\ \emph {et~al.}(2016)\citenamefont
  {Heidenreich}, \citenamefont {Dunkel}, \citenamefont {Klapp},\ and\
  \citenamefont {B{\"a}r}}]{heidenreich2016hydrodynamic}%
  \BibitemOpen
  \bibfield  {author} {\bibinfo {author} {\bibfnamefont {S.}~\bibnamefont
  {Heidenreich}}, \bibinfo {author} {\bibfnamefont {J.}~\bibnamefont {Dunkel}},
  \bibinfo {author} {\bibfnamefont {S.~H.~L.}\ \bibnamefont {Klapp}}, \ and\
  \bibinfo {author} {\bibfnamefont {M.}~\bibnamefont {B{\"a}r}},\ }\href@noop
  {} {\bibfield  {journal} {\bibinfo  {journal} {Phys. Rev. E}\ }\textbf
  {\bibinfo {volume} {94}},\ \bibinfo {pages} {020601} (\bibinfo {year}
  {2016})}\BibitemShut {NoStop}%
\bibitem [{\citenamefont {Reinken}\ \emph {et~al.}(2018)\citenamefont
  {Reinken}, \citenamefont {Klapp}, \citenamefont {B{\"a}r},\ and\
  \citenamefont {Heidenreich}}]{reinken2018derivation}%
  \BibitemOpen
  \bibfield  {author} {\bibinfo {author} {\bibfnamefont {H.}~\bibnamefont
  {Reinken}}, \bibinfo {author} {\bibfnamefont {S.~H.~L.}\ \bibnamefont
  {Klapp}}, \bibinfo {author} {\bibfnamefont {M.}~\bibnamefont {B{\"a}r}}, \
  and\ \bibinfo {author} {\bibfnamefont {S.}~\bibnamefont {Heidenreich}},\
  }\href@noop {} {\bibfield  {journal} {\bibinfo  {journal} {Phys. Rev. E}\
  }\textbf {\bibinfo {volume} {97}},\ \bibinfo {pages} {022613} (\bibinfo
  {year} {2018})}\BibitemShut {NoStop}%
\bibitem [{\citenamefont {Mermin}\ and\ \citenamefont
  {Wagner}(1966)}]{mermin1966absence}%
  \BibitemOpen
  \bibfield  {author} {\bibinfo {author} {\bibfnamefont {N.~D.}\ \bibnamefont
  {Mermin}}\ and\ \bibinfo {author} {\bibfnamefont {H.}~\bibnamefont
  {Wagner}},\ }\href@noop {} {\bibfield  {journal} {\bibinfo  {journal} {Phys.
  Rev. Lett.}\ }\textbf {\bibinfo {volume} {17}},\ \bibinfo {pages} {1133}
  (\bibinfo {year} {1966})}\BibitemShut {NoStop}%
\bibitem [{\citenamefont {Toner}, \citenamefont {Tu},\ and\ \citenamefont
  {Ramaswamy}(2005)}]{toner2005hydrodynamics}%
  \BibitemOpen
  \bibfield  {author} {\bibinfo {author} {\bibfnamefont {J.}~\bibnamefont
  {Toner}}, \bibinfo {author} {\bibfnamefont {Y.}~\bibnamefont {Tu}}, \ and\
  \bibinfo {author} {\bibfnamefont {S.}~\bibnamefont {Ramaswamy}},\ }\href@noop
  {} {\bibfield  {journal} {\bibinfo  {journal} {Ann. Phys. (N. Y.)}\ }\textbf
  {\bibinfo {volume} {318}},\ \bibinfo {pages} {170} (\bibinfo {year}
  {2005})}\BibitemShut {NoStop}%
\bibitem [{\citenamefont {Saintillan}\ and\ \citenamefont
  {Shelley}(2007)}]{saintillan2007orientational}%
  \BibitemOpen
  \bibfield  {author} {\bibinfo {author} {\bibfnamefont {D.}~\bibnamefont
  {Saintillan}}\ and\ \bibinfo {author} {\bibfnamefont {M.~J.}\ \bibnamefont
  {Shelley}},\ }\href@noop {} {\bibfield  {journal} {\bibinfo  {journal} {Phys.
  Rev. Lett.}\ }\textbf {\bibinfo {volume} {99}},\ \bibinfo {pages} {058102}
  (\bibinfo {year} {2007})}\BibitemShut {NoStop}%
\end{thebibliography}%

\end{document}